\documentstyle[12pt]{article}
 
\def\chibar{\overline{\chi}}
\def\psibar{\overline{\psi}}
\def\omegabar{\overline{\omega}}
\def\xibar{\overline{\xi}}
\def\Psibar{\overline{\Psi}}
\def\ssl{s\!\!\!/}     
\def\dslash{\partial\!\!\!/}     

\newcommand{\lsim}{\stackrel{<}{\sim}} 
\newcommand{\nn}{\nonumber}
\newcommand{\ra}{\rightarrow}

\newcommand\be{\begin{equation}}
\newcommand\ee{\end{equation}}
\newcommand\bea{\begin{eqnarray}}
\newcommand\eea{\end{eqnarray}}

\title{                                                                         
{\vspace{-3cm} \normalsize                                                      
\hfill 
\parbox{30mm}{DESY 94-188}
\parbox{30mm}{}}
\\[25mm]                                      
Domain Wall Fermions and Chiral Gauge Theories   \\[8mm]}                     
\author{                                                                        
Karl Jansen \\[4mm]                              
Deutsches Elektronen-Synchrotron DESY, \\                                       
Notkestr.\,85, D-22603 Hamburg, Germany}                                        
                                                                                
\date{October, 1994}                                                             
                                                                                
\begin{document}                                                                
\maketitle                                                                

\begin{abstract}\normalsize

We review the status of 
the domain wall fermion 
approach to construct chiral gauge theories on the lattice.

\end{abstract}

\section{Introduction}

It is a natural question to ask, whether there exists a regularization of 
a chiral gauge theory beyond perturbation
theory. This addresses the pure existence of the Standard Model of 
electroweak interactions.
The astonishing answer to the above question is that a non-perturbative
regularization has so far not been found. Even worse, nogo theorems
\cite{kasmi,nino}
seem to make it impossible to even find such a regularization. 
Of course, the tremendous success of the description of the electroweak
interactions in perturbation theory makes the Standard Model a very
well tested theory in particle physics. 
However, this leaves us with the unsatisfactory situation of the Standard Model
being good ``for all practical purposes''. At this level it is certainly 
not a well founded theory as e.g. QCD. 

The conventional point of view today is to regard the Standard Model
as only an effective theory to describe low energy physics. 
In this low energy limit perturbation theory works very well. However,
due to triviality our theoretical description of the electroweak interactions is
inherently incomplete. At some energy scale which depends on the Higgs boson 
mass the theory has to break down giving room for some --yet unknown-- new
physics. 
This new physics in some sense regulates the minimal
Standard Model. Therefore, finding a non-perturbative regulator for a chiral
gauge theory might give some hint about the theory
``beyond the Standard Model''. 
A non-perturbative chiral regulator would also be important to
understand, whether a construction of an asymptotically free chiral gauge theory
is possible \cite{eipress,jancapri}.
Not surprisingly then, the attempts of a non-perturbative definition of
a chiral gauge theory have been numerous and in particular the search
for lattice chiral fermions has been an intensive area of research in the last
few years. 

The outcome of all attempts to put the Standard Model on the lattice is, 
however,
so far either negative or no convincing evidence
of their success could be given. 
The stumbling block in constructing chiral gauge theories
on the lattice are the famous nogo theorems by Karsten and Smit
\cite{kasmi} and by
Nielsen and Ninomiya \cite{nino}, see also \cite{wilson}. 
The assumptions of the nogo theorems are surprisingly
mild. They require hermiticity, locality, and translational invariance 
of the underlying interaction.
With these assumptions there will always be a clash between chiral gauge 
invariance and the possibility of
a single Weyl fermion on the lattice. 
To each fermion with given chirality there will always be a doubler
mode with the opposite handedness. This proliferation of fermions is deeply
connected to the anomaly structure on the lattice. Either we will have
chiral gauge invariance but then the doubler fermions appear 
and cancel the anomaly, or we lift the
doublers by some mechanism and destroy chiral gauge invariance.
It is worth to note that
these theorems are claimed to hold also without refering to the lattice
at all \cite{nino}. 
This is reminiscent of Adler's demonstration that it is impossible to find
a gauge invariant regularization such that the axial current remains
conserved \cite{adler}. 
It emphasizes the fact, that the question of regularizing chiral fermions
non-perturbatively is not a problem for the lattice alone but one
of principle.

To circumvent the nogo theorems one has then to violate some of the
above assumptions at one stage. 
Although it is acceptable that at some intermediate step one of the above
properties is missing, one clearly would like to recover them in the
continuum theory as they are important ingredients for a sensible theory. 
There are numerous ways trying to find such violations and still recover
the electroweak interaction in the continuum. For an overview I refer to
several review talks in the annual Lattice proceedings \cite{jancapri,
jansei,maatall,istvanjapan,donamster}. 
In particular I would like to draw the reader's attention to the proceedings
of the Rome workshop \cite{rome} which contains 
virtually all approaches at that time. In 
\cite{nielsen}
many of these attempts, the way how they tried to circumvent 
the nogo theorems and reasons why they did not work, are summarized.

One attempt of finding a method to simulate chiral fermions on the lattice
was not covered in the Rome workshop. This was Kaplan's idea 
\cite{david} of generating
a chiral mode on a domain wall that is induced by a mass defect along an
extra dimension. 
Imagine a scenario where we actually live in a 5-dimensional world. 
Assume that along 
the fifth dimension there is a defect generated by a fermion mass term that only
depends on the fifth direction and has the form of a soliton. 
This will generate a domain wall and on this 4-dimensional wall we will
find a chiral zeromode traveling along it \cite{jare,caha}.

In \cite{caha}
Callen and Harvey studied general anomaly descent relations
\cite{anodesc}  connecting
fermion zeromodes on a domain wall 
and anomalies in even with parity anomalies in odd dimensions.
In particular they demonstrated that the odd dimensional 
theory is anomaly free. Performing a Goldstone-Wilczek \cite{gowi}
type of calculation they showed that the 
anomaly in the gauge current due to the chiral zeromode on the domain wall
is exactly cancelled
by the divergence of the Chern-Simons current generated by the heavy fermions
in the fifth dimension off the wall. That the Chern-Simons current is
not divergence free is due to the fact that the mass defect changes sign
when crossing the domain wall.

One interesting aspect of the Callan-Harvey analysis is, that it can be
taken over to the lattice. Indeed, it has been pointed out already in 
\cite{fradkin} 
that this might lead to a realization of chiral lattice fermions
\footnote{
There a domain wall system was studied in the context of a stacking fault in a
PbTe-type crystal. Using the argumentation of Callan and Harvey the authors
of ref.\cite{fradkin} discussed the apparent appearance of the parity anomaly.
Using techniques similar to Callan and Harvey they showed  that the complete
3-dimensional condensed matter system is anomaly free. They proposed that
the domain wall model may be used for solving the chiral fermion problem
in lattice gauge theory. In particular they suggest to introduce the Wilson
mechanism for lifting the doublers. However, they 
then gave up the domain wall model
idea.}.

Independently of \cite{fradkin} Kaplan suggested to use the domain wall
setup to construct a chiral gauge theory on the lattice \cite{david}. 
The hope is to
produce a chiral theory on the 4-dimensional domain wall. Using the 
--in lattice QCD successful--
Wilson mechanism, it can be expected that the dangerous doubler
modes are decoupled.
The difference to other approaches is that the 5-dimensional theory
one would start with is vectorlike from the very beginning thus avoiding
problems with the nogo theorems.
Through the Callan-Harvey mechanism the gauge anomaly would be reproduced
on the domain wall while at the same time the 5-dimensional system
remains anomaly free
due to the mechanism of anomaly cancellation.
Another attractive
property of the domain wall model is, that also the fermion number current is
anomalous on the 4-dimensional wall \cite{david,aoki}, 
although, it is again conserved in the
5-dimensional system. 

This promising picture evokes immediately some questions. Staying in a
situation where the fermions are free or are at most only 
coupled to external gauge fields, 
we have a single Weyl fermion on the wall. Then
how did we circumvent the nogo theorems?
A first answer is that we violated translational invariance. But this is
only true in the 5-dimensional system. With respect to the 4-dimensional
domain wall the system is translational invariant and still we have the
single chiral fermion and the anomalies work out correctly.

The answer to this seemingly paradox can be given, if one looks at the
finite lattice situation. There we have to introduce some kind of boundary
condition in the extra direction. This leads --unavoidingly-- to a second
domain wall on which an additional zeromode appears. This mode has the
opposite chirality as the one on the original wall. Both modes have 
an exponentially
small overlap \cite{david,jansen,janschma}. Therefore we are in full accordance
with the nogo theorems. We indeed have the mode spectrum as demanded by
the theorems. Due to the extra dimension we have just separated them
such that they do not communicate. The whole setup is gauge invariant, 
but we created a loophole for the charges such that they can escape in
the extra dimension leaving a chiral theory on the domain wall.
For the infinite lattice the second domain wall is sent to infinity.
That this picture is indeed correct, as long as we do not have dynamical
gauge fields, could be demonstrated in \cite{jansen,gojan}.

There remains, of course, the crucial point what will happen when 
dynamical gauge fields are added. There have been two main lines of thought.
The first one is to add 5-dimensional gauge fields
giving the gauge fields in the fifth direction a coupling strength
different from the 4-dimensional fields
\cite{david}. A number of papers investigated
this scenario
\cite{yang,prades,prd,aoki,kawano,haoki}. 
The conclusion was that this setup would not lead to the
desired chiral gauge theory. The main reason is that one 
would end up in the so-called 
layered phase where the system reduces to a standard 4-dimensional lattice
system of Wilson fermions and no zeromodes appear. 
Hence the model becomes vectorlike and can
not be used for a construction of a chiral lattice theory.

The alternative approach \cite{david2,prd,nane} 
is to leave the gauge fields strictly 
4-dimensional and to confine it to only one of the walls. 
In this way only one of the domain walls would be gauged and we are left with
a region in the extra direction which is gauged and its ungauged 
complement. For
reasons that will become clear later we call the gauged region the
``waveguide'' region. At the boundary of the waveguide to the ungauged 
part in the fifth direction gauge invariance is broken.
This leads
to a situation where one studies either a gauge non-invariant model or a
model which is gauge invariant but contains scalar fields at
the boundaries of the waveguide. The scalar (or St\"uckelberg) fields serve
to restore gauge invariance in much the same manner as the Standard Model
is made gauge invariant. Adding a scalar field that couples to the fermions
suggests, to also introduce a
Yukawa-coupling $y$. 
It can therefore be expected that fermion masses are generated
via the Higgs mechanism through the Yukawa-coupling of the fermions to the
scalar vacuum expectation value. These fermion modes will live at the
waveguide boundary. 
The crucial question is, whether
these states can be decoupled such that we are left with only the
gauged and ungauged
domain wall zeromodes separated from each other in the fifth dimension.

One may argue
that such a scenario is condemned to fail from the very beginning.
The reason being that approaching the phase transition 
from a spontaneously broken to a symmetric phase the vacuum
expectation value $v$ approaches zero giving rise to small fermion masses which
may be interpreted as mirror fermions. In particular these masses should be
zero in the symmetric phase. Thus they would not decouple. However,
investigations of Yukawa models on the lattice revealed a surprisingly
rich phase structure at large values of the Yukawa-coupling \cite{deje}. 
In particular
a so-called strong coupling region could be identified
\cite{junko,anna,su2pd}. In this region
the Yukawa-coupling becomes so strong that the fermions 
combine with the scalar fields
to form massive bound states with masses at the cut-off. If such a
region would exist also in the domain wall model, the fermions 
at the waveguide boundary
could be made heavy and would decouple.
We then would be left with a single chiral zeromode bound to the domain wall.

Another argumentation which would make the domain wall model fail is that
the charge due to this chiral zeromode will flow off the wall transported
by the Chern-Simons current and would have to be deposited somewhere.
The most easy scenario to imagine is that there are additional fermion
zeromodes that induce a current absorbing this charge
\cite{distler}. However, due to our
waveguide setup, the gauge fields have to stop somewhere in the fifth
dimension. At this point we will have a change of the gauge field that
can induce a Wess-Zumino current. This current can give rise of 
cancelling the anomaly.

From these considerations it is clear, that the waveguide model can not
be ruled out by simple arguments. The question whether the domain wall approach
would be successful or not becomes a dynamical one and a numerical
simulation becomes necessary. Such a simulation for a 3-dimensional
setup was performed in \cite{prd}. There the weak Yukawa-coupling region,
where the fermion masses are proportional to $v$ and therefore vanish in the 
symmetric phase, could be established. For large values of the
Yukawa-coupling the results of this
simulation were compared with Yukawa models that have been
investigated earlier and that do or do not show a strong coupling
behaviour. Strong evidence was collected that in the domain wall model a 
phase with the strong coupling behaviour, as described above, does not exist.
This was confirmed by the
results of a strong coupling expansion which revealed a new phase with
weak coupling behaviour at large $y$. It has to be concluded therefore
that in the domain wall model physics is governed by a weak coupling
behaviour at all values of $y$ thus not leading to a chiral theory.

In this review we will in the following discuss the domain wall model
in more detail.
Throughout this paper the fermions 
are thought to be taken in an anomaly free representation. 
After introducing the fermion doubling problem and how it may
be solved by the domain wall model we will continue to give an explicit
demonstration of the Callan-Harvey picture of charge flow on the lattice.
We then present the model with the scalar fields added. We discuss a large
$y$ expansion in this model and the results of the numerical 
simulations. 
We will present the suggestion that the extra dimension can be held
strictly infinite. We
end with a possible prospect of the domain wall model for
simulating QCD and give a summary and conclusions.

\section{Free fermions, doubling and a possible escape}

In this section 
the problem of fermion doubling on the lattice \cite{janacta} is illustrated 
and the idea of domain wall fermions introduced.
The discussion will be for two dimensions only and the Hamiltonian
formalism will be used. The results obtained in this simple setting 
show already the essential features and are readily
generalized to arbitrary dimensions. In this section we will indicate
the lattice spacing $a$ explicitly in the formulae. Otherwise $a$ is set
to one throughout the paper.
We start our discussion with the 1-dimensional Hamilton operator for
free fermions which --in the continuum-- may be written as
$H_c=-\sigma_1 (\dslash + m_0)$. 
The operator $\dslash =\sigma_2\partial/\partial x$ 
acts only on the $x$ coordinate and the $\sigma$'s are the usual
Pauli matrices. 
$H_c$ describes a single fermion of mass $m_0$ obeying the
relativistic dispersion relation
$E^2 = p^2 +m_0^2$. A ``naive'' discretization of $H_c$ on the 
infinite lattice would 
be
\be \label{eq2.1}
H=-\sigma_1\left[\sigma_2\partial_x + m_0\right]\; ,
\ee
\noindent where $\partial_x$ now denotes the lattice derivative 
\be  \label{eq2.2}
\partial_x =
\frac{1}{2a}\left[\delta_{x,x+\mu} - \delta_{x,x-\mu}\right] ,
\ee
$a$ denotes the lattice spacing and $\mu$ is a displacement by $a$
in a given direction on the lattice. In our 1-dimensional example $\mu =a$.

By Fourier transformation, the particle energies are easily obtained as
\begin{equation}  \label{eq2.3}
E = \pm\sqrt{\frac{1}{a^2}\sin^2 (ak) + m_0^2}\; , 
\end{equation}
with k lying in the Brillouin zone $-\pi < ak \le \pi$. 
For $ka \ll \pi$, the naive continuum limit, 
$a\rightarrow 0$, reproduces the above
relativistic dispersion relation $E^2 = k^2 + m_0^2$. 
However, for momenta $ ak\approx \pi $,
we find 
again $E^2 = k^2 + m_0^2$ and obtain another fermion with mass $m_0$.
Moreover, while the fermion at the origin of the Brillouin zone at
$ak\approx 0$ is a 
right moving particle as can be seen from the group
velocity $dE/dk$, the one appearing at the corner of the Brillouin zone
at $k\approx \pi/a$  is a
leftmover since the $\sin$ function changes its sign there. 
For massless fermions this amounts to find two chiral fermions
with opposite chirality quite in contrast to the single Weyl fermion 
we would have 
started with in the continuum. This is the famous doubling phenomenon for
lattice fermions. The fact that one finds opposite chirality fermions at
the different corners of the Brillouin zone is a consequence of the
``doubler symmetry'' on the lattice \cite{monmue}.
This symmetry transformation is represented by a matrix
\be \label{eq2.4}
M = M_1M_2\; ; \; M_j = i\sigma_3\sigma_j\; .
\ee
If $\psi_k$ is an eigenfunction of the Hamiltonian 
in momentum space, this transformation exchanges the corners of the Brillouin
zone, $\psi_k = M\psi_{k+\pi} = -\sigma_3\psi_{k+\pi}$. Thus, the mode
at the opposite corner of the Brillouin zone has a flipped chirality. The 
doubler symmetry gives in general dimensions an equal number of left and
right handed particles. As the representation of the corresponding
doubler symmetry group is irreducible, doublers have to appear in the 
fermionic lattice spectrum as long as the doubler symmetry is unbroken.
 
The solution of the doubling problem as proposed by Wilson \cite{wilson} is to 
break the doubler symmetry by adding a
second derivative term to the Hamiltonian

\be \label{eq2.5}
H=-\sigma_1\left[\sigma_2\partial_x + m_0-r\Delta_x\right]\; ,
\ee
with 
\be \label{eq2.6}
\Delta_x = \frac{1}{2a}\left[ \delta_{x,x+\mu} + \delta_{x,x-\mu} 
                     - 2\delta_{x,x}
                       \right].
\ee
The spectrum becomes
\be \label{eq2.7}
E^2 = \frac{1}{a^2}\sin^2 (ak) + \left( m_0+\frac{r}{a}(1-\cos (ak))\right)^2\; .
\ee
The mass of the fermions are now given in the $a\ra 0$ limit as
\be \label{eq2.8}
m = m_0 + 2\frac{r}{a}n_\pi
\ee
with $n_\pi =0$ for the origin of the Brillouin zone 
at $ak\approx 0$ and $n_\pi =1$ for the
corner of the Brillouin zone at $ak\approx \pi$.   
Therefore, we obtain 
for the fermion at the origin of the Brillouin zone again the continuum
dispersion relation but the fermions at the corners get an additional piece
proportional to $1/a$. They  will become infinitely heavy as $a\ra 0$
and decouple from the spectrum.
In general dimensions $n_\pi$ counts the number of $\pi's$ at the different
corners of the Brillouin zone. 

Unfortunately, there is a price to pay for getting rid of the extra unwanted
fermions.  
Imagine, we add dynamical
gauge fields to the Hamiltonian (\ref{eq2.5}) by making the lattice derivatives
(\ref{eq2.2}),(\ref{eq2.6}) gauge covariant. 
The Wilson term clearly acts 
as a mass term so that even if we take the fermion mass $m_0$ 
to zero, we still
loose chiral gauge invariance.
Although chiral symmetry can be restored in QCD by carefully tuning the
bare parameters to some critical value, this appears to be a disaster for
the electroweak interactions.  

Indeed, for vectorlike gauge theories as QCD, weak coupling perturbation theory
revealed the restoration of chiral symmetry for the Green functions
in the continuum limit 
\cite{kasmi,boch}. This
opened the road for lattice simulations of QCD because --though 
technically very challenging-- no problem of principle remains. For a
chiral gauge theory using Wilson's approach \cite{janacta,jansei,swift},
weak coupling perturbation theory \cite{hands} showed no success in
obtaining the desired target theory and the doublers could not be 
decoupled. This pessimistic picture was strengthened
by results from lattice simulations which showed that 
for weak coupling the doublers
remain in the spectrum \cite{rome,weakspectrum}. 
Also the hope of decoupling as a non-perturbative
effect failed \cite{gopesm,wolfgang}. 
It seems that in order to obtain a chiral gauge theory
one has to be more ingenious than in QCD.

\subsection{Domain wall fermions}

A possible escape from the doubling problem is due to D. Kaplan
who suggests to send the dangerous
modes into an extra dimension. 
It is known since a long time that there exists a chiral zeromode solution
of the continuum Dirac equation in presence of a soliton
\cite{jare}. 
The idea to use such a kind of solution to construct chiral lattice fermions
is to regard our 4-dimensional world as a domain wall embedded in 5 dimensions.
The interface is induced through a mass defect in form of a soliton.
To stay in our 2-dimensional setup of the previous section we will study
a 3-dimensional system.
Along the third extra dimension denoted by $s$, a mass defect 
is introduced through

\be \label{eq2.9}
m(s) = \left\{ \begin{array}{lll}
                    -m_0; & s\ra -\infty \\
                    +m_0; & s\ra +\infty 
                   \end{array} \right. \;\; .
\ee
The concrete form --aside that it has to be monotonic-- 
of the mass is not very important. 
It is easy to show that in this 
situation there exist solutions 
$\Psi_\pm$ which are energy eigenstates of the continuum
Hamiltonian 
\be \label{eq.2.10}
H_c = -\sigma_1\left[\sigma_2\partial(x) + \sigma_3\partial(s) + m(s)\right]\; , 
\ee
\be \label{eq.2.11}
\Psi_\pm = e^{ip_x x} \Phi_\pm(s) u_\pm
\ee
with
\be \label{eq2.12}
\Phi_\pm(s) = \exp\left(\pm\int_0^s m(s') ds'\right)
\ee
and $u_\pm$ a chiral eigenstate, $\sigma_3 u_\pm =\pm u_\pm$.
Only the function $\Phi_{-}(s)$ corresponds to a {\em normalizable}
solution. 
This solution describes a
chiral zeromode traveling along the interface. 
It is bound to the wall and falls off
exponentially with growing distance from the wall. 

Translating this model to the --infinite-- lattice we 
choose a step function for the
mass 
\be \label{eq2.13}
m(s) = m_0 \theta(s);\;\;\theta(s)=\left\{ \begin{array}{lll}
                    -1; & s\le -a \\
                     0; & s = 0\\
                    +1; & s\ge a  
                   \end{array} \right. \;\; .
\ee
The continuum derivatives are again replaced by finite lattice differences
and the Hamiltonian reads now
\be \label{eq2.14}
H=-\sigma_1\left[\sigma_2\partial_x + \sigma_3 \partial_s + m(s)\right]\; .
\ee
Imposing an ansatz similar to the continuum solution
\be \label{eq2.15}
\Psi_\pm = e^{ikx}\Phi_\pm(s) u_\pm
\ee
we find again a normalizable solution 
\be \label{eq2.16}
\Phi_{-}(s) = e^{-\mu_0 |s| }\;\; ,
\ee
with $\sinh\mu_0 =m_0$.
This is a solution that is bound to the domain wall, it is chiral and
describes plain waves along the direction of the domain wall. Unfortunately,
there exists another solution which only appears on the lattice 
\be \label{eq2.17}
\Psi_{+} = e^{ikx}\Phi_{+}u_{+}
\ee
with 
\be \label{eq2.18} 
\Phi_{+}(s) = (-1)^{s} \Phi_{-}(s)\; .
\ee
This solution describes again a massless fermion traveling 
along the domain wall,
it is also bound to it and has opposite chirality
as compared to the solution $\Phi_{-}(s)$, eq.(\ref{eq2.16}). 
In other words it is a doubler fermion
in the s-direction! Even worse, also the doubler mode in the $x$-direction
from the corner
of the Brillouin zone is a solution.
Therefore the spectrum on the domain wall consists of
two chiral fermions with positive chirality and their two doublers --one in the
$s$ and one in the $x$ direction-- with
opposite chirality.

What we have achieved so far is instead of removing the doubler modes we
have created one in addition. But what if we now try to apply Wilson's trick
adding a higher derivative term? 
The 3-dimensional model we started with is
vectorlike as in QCD and we saw that there the Wilson term is harmless.
The Hamiltonian becomes
\be \label{eq2.19}
H=-\sigma_1\left[\sigma_2\partial_x + \sigma_3 \partial_s + m(s)
- r(\Delta_x + \Delta_s) \right]\; .
\ee
Let us set the Wilson coupling $r=1$ for the moment and
again try an exponential ansatz for the transverse wavefunction $\Phi$ keeping
plane wave solutions in the $x$-direction
\be \label{eq2.20}
\Phi_\pm(s\pm a) = - k_{eff}(s) \Phi_\pm(s)
\ee
with an effective momentum
\be \label{eq2.21}
k_{eff} (s) = m_0\theta(s) -1 - F(k);\;\; F(k) = 1 - \cos(ak)\;\; .
\ee
In order to obtain {\em normalizable} solutions,  
the functions $\Phi$ have to decrease with growing $|s|$ on both sides of the
wall. This
implies the following conditions
to be fulfilled
\bea \label{eq2.22}
\begin{array}{cccccccc}
\Phi_+: & |k_{eff}| > 1 & \mbox{for} & s < 0, & |k_{eff}| < 1 & s > 0 \\
\Phi_-: & |k_{eff}| < 1 & \mbox{for} & s < 0, & |k_{eff}| > 1 & s > 0\; .
\end{array}
\eea
One recognizes that only one of the solutions is normalizable
namely $\Phi_-$ and that we have to throw away the $\Phi_+$ solution.
Thus we are left with only one zeromode solution and we got rid of the doubler
mode in the $s$-direction (\ref{eq2.18}).
Since the
doubler mode in the $x$-direction is decoupled by the usual Wilson
mechanism we remain
with a single chiral fermion solution as in the continuum.
This chiral zeromode is bound to the domain wall and exponentially
decreases going away from the wall.
The normalizability condition (\ref{eq2.22}) requires
\be \label{eq2.23}
0 < m_0 - F(k) < 2\;\; .
\ee
This means that for a given value of the domain wall height $m_0$ the
chiral zeromode does not exist for all momenta but only up to some
critical momentum $k_c$. At this value the fermion ceases to be chiral and
vanishes in a band of heavy modes. This behaviour is the important ingredient
of the domain wall model. The chiral zeromode only exists as a low
energy phenomenon. For large values of the momentum it becomes heavy and
is not distinguishable from the doubler modes. It therefore does not appear
again as a low momentum state at some other corner of the Brillouin zone.

In \cite{janschma} the zeromode solutions for general $m_0$ and $r$ values 
were computed. 
There the 3-di\-men\-si\-o\-nal Dirac-operator was studied. Therefore the momentum
space becomes 2-di\-men\-si\-o\-nal.
The normalizability condition led to a very peculiar
zeromode spectrum (see also fig.1 in \cite{janschma}). 
For $m/r \le0$ no chiral fermions exist. Increasing $m/r$ a chiral
fermion is found with the critical momentum given by $m_0 = rF(k)$. Increasing
$m/r$ further the value of the critical momentum keeps growing, too. This
continues until at $m/r =2$ the maximal critical momentum is reached.
For $m/r > 2$ the chiral zeromode at the origin of the Brillouin zone 
$\vec{k}=(0,0)$ is
lost. Instead new zeromodes appear at the corners of the Brillouin zone
$\vec{k}=(\pi,0)$ and $\vec{k}=(0,\pi)$.
They have the opposite chirality as the one at $k\approx 0$. The critical
momenta of these new zeromodes are in the interval determined by $m_0 =rF(k)$
for the lower and $m_0 = r(F(k)+2)$ for the upper momentum
interval boundary. Increasing $m/r$ even further we loose these modes again
for $m/r > 4$ but get a new zeromode at $\vec{k}=(\pi,\pi)$ with flipped 
chirality. Finally, for $m/r > 6$ the chiral zeromodes disappear completely. 
Depending on the bare parameters of the domain wall model one can therefore
generate a chiral zeromode spectrum as needed.

\subsection{The domain wall model on the finite lattice}

The interesting question is, of course, whether the above scenario
 can
be realized on a finite lattice. There one has the additional complication
that some kind of boundary condition has to be imposed which necessarily 
induces a second domain wall. 
In the following we let the lattice have finite
extensions $L_s$ in the $s$ and $L$ in the $x$-direction and have
$s\in [1,L_s]$, $x\in[1,L]$.
The domain wall mass will be chosen 
\be \label{eq2.24}
m(s) \equiv \sinh(\mu_0)\theta(s) ;\;\;
\theta(s) = \left\{ \begin{array}{lll}
                    -1 & 2\le s \le \frac{L_s}{2} \\
                    +1 & \frac{L_s}{2}+2 \le s \le L_s \\
                     0 & s=1, \frac{L_s}{2}+1
                     \end{array} \right. \;\; .
\ee
The finite lattice Hamiltonian including the Wilson term reads

\begin{equation} \label{eq2.25}
H  =-\sigma_1\left[\sigma_2\partial_x + \sigma_3\partial_s 
+m(s)-r(\Delta_x +\Delta_s) \right] \;\; .  
\end{equation}
Due to the finite lattice extension, we have to specify some sort of boundary
condition which will be chosen to be periodic in the $s$ and antiperiodic
in the $x$-direction. 
The Hamiltonian (\ref{eq2.25}) can be reduced to only depend on
$s$ by imposing plane wave solutions in the $x$-direction. One obtains,
setting $a=1$,

\begin{equation} \label{eq2.26}
H  =-\sigma_1\left[\sigma_2\sin(k) + \sigma_3\partial_s 
+m(s)+r(\cos(k)-1) -r\Delta_s) \right] \;\; ,
\end{equation}
where the momenta $k$ are now discretized, 
\be \label{eq2.27}
k=\frac{\pi}{L}(n+\frac{1}{2})\; , \; n=0,...,L-1\; .
\ee
This choice of the momenta $k$ corresponds to antiperiodic boundary 
conditions which will also be chosen in numerical investigations 
presented later. They become necessary in numerical simulations
to avoid exact zeromodes which render the simulation algorithms
impractical.

In this form the energy eigenvalues and the corresponding eigenfunctions
for a given value of $k$
can be obtained by diagonalizing the Hamiltonian which 
is a $2L_s\otimes 2L_s$ matrix numerically. In this way one gets the
momentum dependent 
eigenvalues $\lambda_{\pm k}$ and their corresponding wavefunctions. 
The eigenvalues come in $\pm$ pairs. In fig.1a the
wavefunctions belonging to the lowest positive eigenvalue
$\lambda_0$ and its
negative partner are plotted.
One clearly sees that the second domain wall generates an additional solution
which is absent on the infinite
lattice. The nice thing is that both solutions are located on each of their
domain walls at $s=1$ and $s=L_s/2 +1$. They have opposite chirality
and fall off exponentially 
going away from the domain wall having only an exponentially small overlap
$\propto e^{-\mu_0L_s}$. 

\begin{figure}                                                              
                                                                              
\vspace{10cm}
\includegraphics{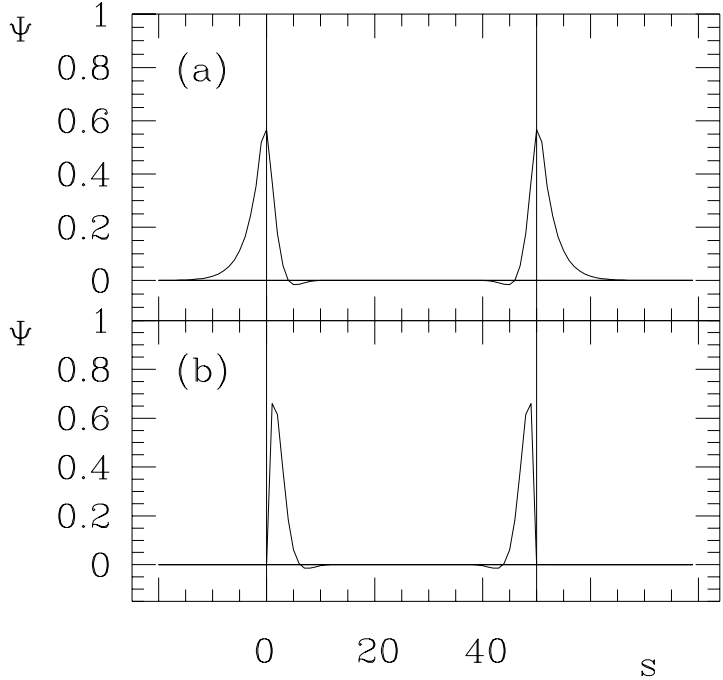}
\caption{ \label{fig1}                                                         
The zeromode wavefunctions as obtained from the 
finite lattice Hamiltonian (\protect{\ref{eq2.26}}). 
The domain wall
mass $\mu_0 =1$, the Wilson coupling $r=1$ and the lattice size in the
$x$-direction is $L=32$. The horizontal lines indicate the position of
the domain walls. Fig.~1a is the domain wall model with an extension of $L_s =100$
in the extra dimension. Fig.~1b is the variant of this model with open
boundary conditions and an extension of $L_s =50$ only. Both figures show
that the zeromode wavefunctions are sharply localized on the walls and fall
off very rapidly with growing distance from the wall. 
}                                                                              
\end{figure}                                                                   

In fig.2a the lowest and the next to lowest positive 
eigenvalues are plotted as a 
function of the lattice momenta k
corresponding to antiperiodic boundary conditions. 
One observes the existence of
the critical momentum $k_c$. The lowest eigenvalue $\lambda_0$ follows
the dispersion relation of a free lattice fermion $\sin(k)$ up to the 
critical momentum where it vanishes in a band of larger eigenvalues.
In fig.~2a also the next lowest positive eigenvalue $\lambda_1$ is
plotted. For low momenta where $\lambda_0 (k) = \sin(k)$ it stays
at the order of the lattice cut-off exhibiting a large gap to the
lowest eigenvalue. It stays almost constant as a function of the
momenta until at $k_c \approx 1$ it combines with $\lambda_0$.
Fig.~2b shows that
the mode corresponding to the lowest positive
eigenvalue really is a chiral fermion.
The ratio build from the wavefunction $\psi_0$ belonging to $\lambda_0$
\be \label{eq2.28}
R=\frac{<\psibar_0\psi_0 >}{<\psibar_0\sigma_1\psi_0 >}
\ee
indicates the chirality of a mode. It is zero when $\psi$ is a chiral eigenstate
and becomes non-zero if chirality is lost. The figure shows that this happens
exactly at the place where the gap $\lambda_1 - \lambda_0$ in fig.2a shrinks
to zero which is at $k_c$.

\begin{figure}                                                              
                                                                              
\vspace{10cm}
\includegraphics{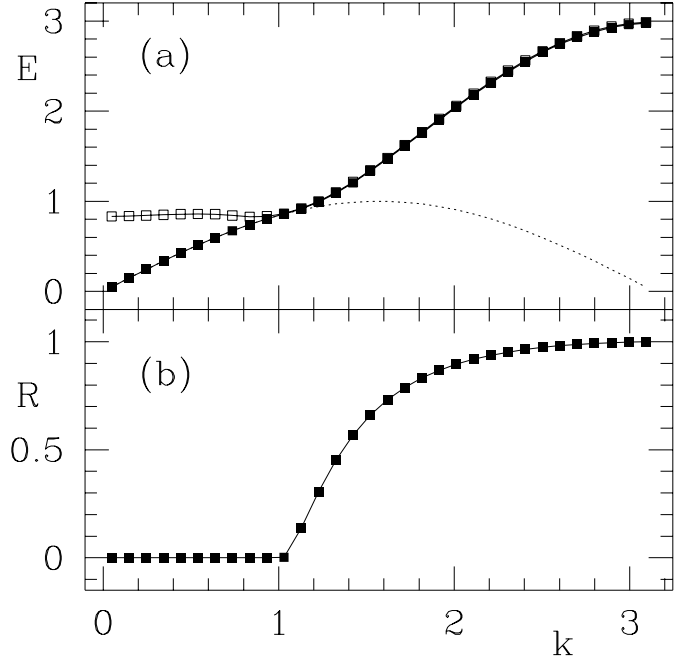}
\caption{ \label{fig2}                                                         
The lowest and next to lowest positive eigenvalues are 
plotted in fig.2a against the
lattice momenta $k=\pi(n+0.5)/L$ with $n=0,...,31$ corresponding to a
lattice of 32 points in the x-direction with anti periodic
boundary conditions. The parameters are the same as in 
fig.1. The dotted line is the energy of a free massless fermion on the
lattice, $\sin(k)$. One nicely observes that the lowest eigenvalue $\lambda_0$
(full squares) 
follows the $\sin(k)$ behaviour whereas the next mode 
$\lambda_1$ (open squares) is at the order
of the cutoff and stays almost constant as a function of $k$. At a critical 
value of the momentum $k_c\approx 1$ the gap $\lambda_1 - \lambda_0$ shrinks to
zero. At this point the lowest mode departures from the sin behaviour.
Fig.~2b shows that the mode corresponding to $\lambda_0$ is really chiral.
Plotted is the ratio $R=<\psibar\psi>/<\psibar\sigma_1\psi >$ for the
eigenfunction of $\lambda_0$. $R$ is zero if $\psi$ is a chiral eigenfunction
and deviates from zero if the chirality is lost. $R$ becomes non-zero at exactly
the value of $k$ where the two modes in fig.2a combine, i.e. at $k_c$.
}                                                                              
\end{figure}                                                                   

\section{The anomaly}

The fermion spectrum we have obtained in the previous section looks very
promising.
We have two Weyl fermions on the
finite lattice living on each of the domain walls. They
are separated in the extra dimension and have an only
exponentially small overlap. Besides the chiral fermions we have the heavy
doubler modes on the walls and additional heavy fermions off the wall
with masses at the order of the cut-off.
For free domain wall fermions 
the low energy physics on both walls appears therefore to describe a chiral
theory.

Let us assume that adding gauge fields will not destroy this picture.
This is, of course, a strong assumption but, as we will see, it is at
least justified for weak {\em external} gauge fields. Introducing
{\em dynamical} gauge fields is much more problematic and will be discussed
later. 
A (2-dimensional) physicist living on one of the walls knowing 
nothing about he extra dimension would eventually discover 
``his'' Standard Model
-- the 2-dimensional chiral Schwinger model.
He would also discover the chiral anomaly and wonder
how he could regularize his chiral theory. If the domain wall model is able 
to succeed in producing a chiral gauge theory on a wall as the low energy
limit of the 3-dimensional model, it should as a first step generate
the correct anomaly structure for external gauge fields. Therefore the
computation of the anomaly in the 2-dimensional Schwinger model
is a first crucial step in testing the domain wall approach. 
Since at the present stage the gauge fields are only external, we might
select our physical wall by hand. It is therefore irrelevant that
the gauge fields connect both chiral zeromodes. What we are aiming at, is
to get a picture of the charge flow in the 
finite lattice model containing both walls.
For the
following calculations we 
use a variant of the domain wall model as introduced by Shamir 
\cite{shamir93} and
follow \cite{creutz} to compute the axial
charge in the Hamiltonian formalism from the spectrum flow. 

Shamir's variant is to not let the mass change its sign across the second 
domain wall but to cut it off completely and introduce open boundary conditions
instead. The system can be regarded as being in a box with infinitely
high walls at the ends of the extra dimension. 
The Hamiltonian analysis of this situation 
is very similar to the above discussion
\cite{shamir93}. 
In this setup the chiral zeromodes
appear as surface modes on the walls with opposite chirality on the two 
borders. 
We show in fig.1b the zeromode spectrum of the Hamilton
operator with open boundary conditions in the $s$-direction.
As before we keep 
antiperiodic boundary conditions in the $x$-direction and choose the same
parameters as in fig.1a.
The figure shows the
surface modes which are again separated in the
extra dimension. They have only exponentially small overlaps and are both
eigenstates of $\sigma_3$, the chiral operator.
It is remarkable how similar they are in shape compared to the domain wall
zeromodes depicted in fig.1a. The boundary version of the domain wall model
has the advantage that the lattice in the $s$-direction is only half that of
the domain wall model. This might lead to an improvement in practice if one
thinks of numerical simulations. This would be the case in particular when
the domain wall fermions are going to be used for simulations of QCD, a prospect
we will discuss later.

To see that there really flows an anomalous current along the domain wall
we apply a time dependent external field in the $x$-direction. 
We introduce the
gauge covariant lattice derivatives
\be \label{eq3.1}
\partial_x(U) = \frac{1}{2}\left[U_{x,\mu}\delta_{x,x+\mu} 
              - U_{x-\mu,\mu}^{*}\delta_{x-\mu,x}\right]
\ee
and
\be \label{eq3.2}
\Delta_x(U) = \frac{1}{2}\left[ U_{x,\mu}\delta_{x,x+\mu} 
                + U_{x-\mu,\mu}^{*}\delta_{x-\mu,x} - 2\delta{x,x}
       \right]\; .
\ee
Here $U$ denotes the usual compact representation of the gauge field. In this
review we will for all discussions only use $U(1)$ gauge fields. Then $U$ is
a phase, $U_{x,\mu} = e^{\{i\alpha_{x,\mu}\}}$ and is related to the
continuum gauge potential $A_\mu(x)$ by $e^{iqA_\mu(x)}$. 
The gauge fields will be chosen to be only 2-dimensional and identical
on every $s$-slice. Although this would lead to an interaction of the
surface modes when the gauge fields were fully dynamical, this is not 
a problem for only external gauge fields because in this case a domain wall
can be singled out by hand. 

The Hamilton operator becomes
\be \label{eq3.3}
H=-\sigma_1\left[\sigma_2\partial_x(U) + \sigma_3 \partial_s + m(s)
- r(\Delta_x(U) + \Delta_s) \right]\; .
\ee
If we think of $H$ being written in terms of creation and annihilation
operators $\hat{\psi}^\dagger$ and $\hat{\psi}$, it is 
gauge invariant under a gauge transformation
\be \label{eq3.4}
U_{x,\mu} \rightarrow g_xU_{x,\mu}g_{x+\mu}^*\; ; \psi_x\rightarrow g_x\psi_x,
\; g\in U(1)\; .
\ee
We will choose an external field that is constant in space and varies only
in time. For adiabatic external fields we expect then from the
continuum
\bea \label{eq3.5}
 Q_3(t) & \equiv & \frac{q}{2\pi}\int_0^tdt'\int dx \epsilon_{\mu\nu}
                    F^{\mu\nu}(x,t')\nn \\
              & = & -\frac{q}{2\pi} L\int_0^t 2dt'\frac{\partial}{\partial t'}
                  A(x,t') \nn \\
              & = & -2q\frac{L}{2\pi} A_1(t)\; .
\eea
Here we identify the gauge potential $A_1(t) = q\alpha(t)$, with $q=1$ the
gauge charge and $Q_3$ 
denotes the (axial) charge. 
Due to the gauge invariance (\ref{eq3.4}) the Hamiltonian (\ref{eq3.3}) 
depends only on the product of all the gauge links. Thus
the gauge potential 
$U=e^{i\alpha}$ becomes a constant global phase factor 
acquired by a fermion traveling around the finite lattice system. The
lattice Hamiltonian reads in momentum space 
\begin{equation} \label{eq3.6}
H  =-\sigma_1\left[\sigma_2\sin(k-\alpha) + \sigma_3\partial_s 
+m(s)+r(\cos(k-\alpha)-1) -r\Delta_s) \right] \;\; .  
\end{equation}
This is the 
free Hamiltonian with momenta shifted by the phase $\alpha$.

A definition of the axial charge is quite arbitrary on the lattice 
\cite{shamir93.2}. Basically one has to fulfill that opposite charge
has to be assigned to the two chiral modes.
Defining the vacuum as the state with all negative energy levels filled,
we can define a measure of the charge by
\be \label{eq3.7}
Q_3 =\frac{1}{L_s-1} \sum_{q,s} (L_s -1 -2s) \psi_{s,q}^\dagger\psi_{q,s}
\ee
where the sum goes over the one particle 
wavefunctions $\psi_{s,q}$ above the vacuum. The definition
of $Q_3$ in (\ref{eq3.7}) has the property to be one, if the mode is
located exactly to the left domain wall and minus one if it is located
exactly to the opposite wall. For heavy modes that are smeared over the lattice
the contribution to $Q_3$ is zero. In \cite{creutz} the charge $Q_3$ was 
measured. The result is shown in fig.3 where we plot the charge $Q_3$ as 
a function of the flux $\alpha$. 
If we measure $\alpha$ in units of $2\pi/L$
the anomaly equation (\ref{eq3.5}) becomes $Q_3 = -2\alpha$.
This behaviour can indeed be observed in fig.3.

If the external field would really be adiabatic, the turning on of
the field would be so slow that the surface modes on the walls can tunnel
through the extra dimension, exchanging there roles. This would happen at
$\alpha=1/2$. At this point we will have a change of the
zeromode spectrum (see section 2.1). 
A filled level on one of the boundaries is just about
to become the positive energy particle and an empty level is about to drop
in the sea. We would observe a jump in the charge by $-2$
and the system would relax to the ground state at $\alpha=1$ resulting
in a zero net charge. This is indicated
in fig.3 as the solid line.

We saw, however, that the wavefunctions of the surface fermions are sharply
peaked at the boundary, see fig.1b. Therefore their overlap is exponentially
small resulting in a very tiny tunnel energy $\delta$. For the situation
in fig.3, $\delta \approx 10^{-9}$. Thus $1/\delta$ is the largest timescale 
in the problem. Instead of tunneling of both surface states, a particle-hole
pair is created and the charge keeps decreasing towards $Q_3 = -2$
at $\alpha =1$. This is represented as the stars in fig.3. 

\begin{figure}                                                              
                                                                              
\vspace{10cm}
\includegraphics{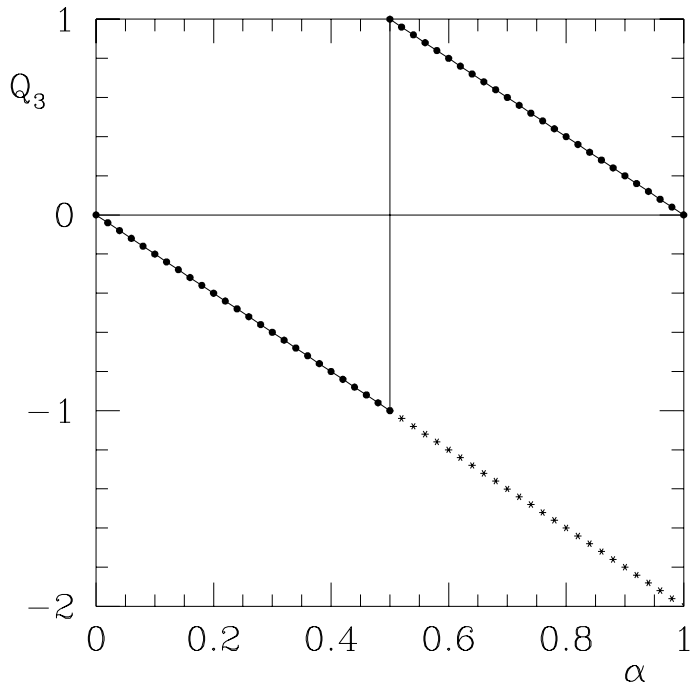}
\caption{ \label{fig3}                                                         
The charge $Q_3$ as function of the ``flux'' $\alpha$. The slope is 
$\approx -2$ as
expected from the continuum anomaly equation. 
}                                                                              
\end{figure}                                                                   

\subsection{The resolution of a puzzle}

For the reader who
knows of the extra dimension there appears to be a puzzle from the results
of the previous section.
In every step of the above discussion the odd dimensional model was gauge
invariant and vectorlike. We are also in full accordance with the nogo
theorems in that there are an equal number of left and right handed fermions.
The only trick to circumvent them is that these modes are separated in the
extra dimension. In fact, the whole idea was to study a vector theory in
order to be able to use the Wilson mechanism in close analogy to lattice
QCD to decouple the doublers.

So, how can there be then anomalous charge creation? The resolution of this
puzzle are the anomaly descent relations
\cite{caha,anodesc}. Callan and Harvey pointed out
that in an odd dimensional theory with a mass defect in form of a domain
wall the heavy modes residing in the extra dimension
off the domain wall do not entirely decouple but produce
a Chern-Simons term in the effective action when they are
integrated out. This action generates a Chern-Simons
current which is very similar to a Hall-current in solid state
Physics. It flows along the extra dimension and perpendicular to the
gauge current induced by the external field. The Chern-Simons current is
responsible for a charge transport in the extra dimension from one of
the walls to the other. 
The strength of the Chern-Simons current 
is different
on the two sides of the wall. Therefore the Chern-Simons current is
not divergenceless. Indeed, as could be shown by Callan and Harvey, the 
divergence of this current is exactly equal in strength to the 
anomaly on the wall. 
This mechanism makes the even
dimensional observer think that charge just ``disappears'' or is
``generated'' out of nowhere.
The total 3-dimensional system is anomaly free
and contains an equal number of left and right handed fermions just as
demanded by the nogo theorems.

With these remarks we leave now the Hamiltonian formalism to discuss the
domain wall model and go to the euclidean action formulation in which
simulations have to be performed in the end. We start with an explicit
lattice computation of the Chern-Simons current, confirming the
continuum picture of the anomaly cancellation by 
Callan and Harvey being realized on the lattice, too.

\section{Euclidean lattice action for domain wall fermions}

The Hamilton formalism for free domain wall fermions
discussed in the previous two sections led to some
remarkable properties of the domain wall model
in its wall-anti-wall realization on a finite lattice. There is a single Weyl
fermion bound to the domain wall with 
its chiral partner living on the anti domain wall separated through
the extra dimension. All other doubler modes are heavy and decoupled by means
of the Wilson mechanism. From the spectral flow we saw that the charge on the 
domain wall very closely followed the continuum anomaly equation.
These properties are certainly very essential building blocks should the
model succeed in describing a chiral gauge theory on the lattice.
Would it have failed at this point, there would have been no reason to continue
with its investigation. The considerations of the last sections can be
regarded as some necessary first checks every lattice chiral fermion
proposal should pass in order to have a chance of being successful. 

Our aim is a non-perturbative understanding of a chiral lattice model.
This will most probably involve a numerical simulation at one stage.
Such simulations are done by evaluating the --euclidean-- path integral
\be \label{eq4.1}
Z = \int{\cal D}\psi{\cal D}\overline{\psi}{\cal D} U e^{S_G + S_F}
\ee
with $S_G$ the gauge field action and $S_F$ the fermionic part.
We will not give the explicit form of both actions. Various alternatives
will be discussed in the next section when possible additions of dynamical
gauge fields to the domain wall model are discussed.
Since we are aiming at a euclidean lattice description,
we switch now the language going from the Hamiltonian formalism to
the path integral formulation of 
the model. We start by performing an explicit calculation of the
Chern-Simons current on the lattice. We will in the following stay in the 
$d=2+1$ dimensional setting and will always have abelian gauge fields.

In their continuum calculation,
Callan and Harvey pointed out \cite{caha} that in a situation
with a mass defect in odd dimensions there is a Chern-Simons current which
is generated by the heavy fermions living far off the wall. This current
has opposite sign on the two sides of the wall and its divergence is hence
non-zero. In fact, Callan and Harvey computed the divergence of the Chern-Simons
current and found that it exactly reproduces the coefficient of the 
anomaly generated by the chiral zeromode on the domain wall. It is 
this mechanism which gives hope that the domain wall model is able to
provide the anomalies which are needed for a reproduction of the anomaly
structure in the Standard Model.

The Callan-Harvey analysis can be considered to be incomplete in two aspects.
($i$) The computation of Callan and Harvey has been done far off the wall
and a computation also close to the wall would be desirable. 
($ii$) The Chern-Simons
coefficient is regulator dependent and it should be computed in 
different regularizations. 
In fact, the continuum calculation of Callan and Harvey was done without 
imposing a regularization at all.
We will improve this 
situation by presenting an analysis of the Chern-Simons current that is valid
also arbitrary close to the wall \cite{chandra} and give an explicit
lattice computation of the Chern-Simons current \cite{colue,gojan,so}.
 
($i$) The effective Chern-Simons action is obtained 
when the heavy fermion modes 
are integrated out. The gauge variation of this action then leads to the
Chern-Simons current the coefficient of which can be calculated
in the low energy limit. This program has been gone through in the
continuum by Callan and Harvey.
However, their computation was done far off the domain wall where the mass
defect can be considered as constant. It would be obviously
desirable to have an analogous calculation also arbitrary close to the wall
\footnote{
This is even more so for the following reason. In \cite{naculich} it was pointed
out that on a domain wall one should find the covariant anomaly. This is
indicated by the strength of divergence of the Chern-Simons current. On the
other hand, the anomaly of a 2-dimensional chiral theory is the consistent
one as required by the Wess-Zumino consistency condition \cite{wess}.
In \cite{bardeen}  it was shown that by adding an extra term the
consistent anomaly can be made covariant. It may be expected \cite{naculich}
that such an extra piece can be provided in the 2-dimensional system 
from the effective Chern-Simons action in 3-dimensions.}.

Such a continuum calculation has 
recently been given in \cite{chandra}.
Choosing the mass defect in the form
\be \label{eq4.2}
m(s) = m_0\tanh (m_0 s)
\ee
we search for solutions of the ei\-gen\-value e\-qua\-ti\-on in Min\-kows\-ki 
space 
$\sigma_1\left\{i\sigma^\mu\partial/\partial x_\mu +m(s)\right\}
\psi_\lambda = \lambda\psi_\lambda$, $\mu=1,2,3$.  
The problem of solving this equation for
the 3-dimensional free domain wall system can be transformed to 
that of a quantum
mechanical scattering problem with a modified 
P\"oschl-Teller potential which can be solved exactly. 
Thus we not only get the chiral
zeromode solution discussed in section 2, but also all the heavy fermion
solutions. From the wavefunctions $\psi_\lambda$ and the eigenvalues
$\lambda$ the propagator $G(z,z')$, $z=(x,t,s)$ can be calculated
\be \label{eq4.3}
G(z,z') = \sum_\lambda \frac{\psi_\lambda(z)\psibar_\lambda(z')}{\lambda}\; .
\ee
In momentum space $G$ is found to consist of two parts, 
$G=G_{chiral}+G_{massive}$.
For $G_{chiral}$ one finds 
\be \label{eq4.4}
G_{chiral}(z,z') \propto \int\frac{d^2 k}{(2\pi)^2}\frac{\sigma_1k_1 + 
                     \sigma_2k_2}
                      {k_0^2-k_1^2}e^{-i(x-x')k_1 - i(t-t')k_2}\; .
\ee
This is the 2-dimensional massless propagator of a chiral fermion.
The prefactor which is left out in (\ref{eq4.4}) gives the $s$-dependence
of the propagator. It reproduces the result
we found earlier that 
it is bound to the wall with an exponential fall off when going away. 
A similar form for the propagator exhibiting its chiral structure was obtained 
in \cite{shamir92,aoki}. In addition to the chiral propagator 
also the one for the heavy modes
can be computed
\be \label{eq4.5}
G_{massive}(z,z') = \int\frac{d^3 k}{(2\pi)^3} \frac{\sigma^\mu k_\mu +M}
              {k^2 -m_0^2}e^{-i(z-z')k}\; ;\mu=1,2,3
\ee
with 
\be \label{eq4.6} 
M= \left( \begin{array}{cc}
            -m(s) & m_{12}(s) \\
              0   & -m(s')
             \end{array}\right)
\ee
where
\be \label{eq4.7}
  m_{12} = \frac{k_0 +k_1}{k_2^2 + m_0^2}
             \left[m(s)m(s') +ik_2(m(s') -m(s)) -m_0^2\right].
\ee
This describes a massive fermion.
By analytic continuation to imaginary time the propagators 
(\ref{eq4.4}) and (\ref{eq4.5}) can
be written in euclidean space where they will be denoted by $S_E$ 
and can such be used for the computation
of the effective action in background gauge fields $A_\mu$ which in 1-loop is
\be \label{eq4.8}
S_{eff}[A] = \frac{1}{2}\int d^3zd^3z' A_\mu(z)V^{\mu\nu}A_\nu(z')
\ee
where
\be \label{eq4.9}
V^{\mu\nu} = \mbox{Tr} \gamma^\mu S_E(z,z')\gamma^\nu S_E(z,z')\; .
\ee

The expression in (\ref{eq4.8}) can be calculated in the $m_0\rightarrow
\infty$ limit taking the gauge fields to be slowly varying. One finds the
effective action to consist of two parts
\be \label{eq4.10}
S_{eff} = S_{eff}^{CS} + S_{eff}^{chiral}
\ee
with a Chern-Simons piece
\be \label{eq4.11}
S_{eff}^{CS} = \epsilon^{\mu\rho\nu}\frac{-i}{8\pi}\int d^3z\mbox{sign}(s)
               A_\mu\partial_\rho A_\nu
\ee
with $\mu ,\nu, \rho =1,2,3$ and a chiral piece
\be \label{eq4.12}
S_{eff}^{chiral} = -\frac{m_0^2}{32\pi^2}\int d^3zd^3z'
                   A_a(z)A_b(z') \frac{y^ay^{*b} + y^by^{*a}}{\|y\|^4}
                   \left[1-f(y,s-s')\right]^2
                  \mbox{sech}^2(m_0s)\mbox{sech}^2(m_0s')
\ee
where $y=(x-x',t-t')$, $y^{a*} = i\epsilon^a_by_b$ is the dual of $y^a$, the
2-dimensional index $a = x,t$ and
\be \label{eq4.13}
f(y,s-s') = \frac{e^{-m_0\|z\|}}{\|z\|}\left[ \|z\|\cosh(m_0(s-s'))
          + (s-s') \sinh(m_0(s-s'))\right]\; .
\ee

The Chern-Simons part $S_{eff}^{CS}$ is the known result \cite{caha}. We
see, however, that there appears an additional piece $S_{eff}^{chiral}$
which has not been computed before. As we will see now this piece is the
extra part as advocated in \cite{naculich,bardeen} to find the covariant
anomaly on the domain wall. We compute the currents by a gauge variation
\be \label{eq4.14}
J_\mu = \frac{\delta S_{eff}}{\delta A_\mu}
\ee
\be  \label{eq4.15}
J_\mu = -\frac{i}{4\pi}\mbox{sign}(s)\epsilon_{\mu\rho\nu}\partial^\rho A^\mu
        + \frac{i}{4\pi} \delta_{\mu a}\epsilon_{ab}A^b\delta(s)\; .
\ee
From the current in (\ref{eq4.15}) we can see that we have obtained the
Chern-Simons current as the first term. Its divergence comes from the
different sign of the mass on the two sides of the wall. There is
an additional piece in the current, residing exactly on the wall. 
It is this piece that
gives an extra contribution to the divergence equations, making the anomaly
on the wall the covariant one. 

($ii$) We now proceed, discussing the second criticism of the
Callan-Harvey computation, the regularization dependence of the 
Chern-Simons coefficient. For this purpose
we describe a computation of the Chern-Simons current 
imposing a lattice regularization. We will work on an infinite lattice and
keep the odd dimensions $d=3$, although the results hold for arbitrary odd 
dimensions.
We are interested in the low 
energy coefficient $c$ of the Chern-Simons
action $\Gamma_{CS}$ which is obtained when the heavy fermions 
are integrated out. This leads to an effective action $S_{eff} = c\Gamma_{CS}$,
\be \label{eq4.16}
\Gamma_{CS} = \epsilon_{\mu\nu\rho}\int d^3 x A_\mu \partial_\nu A_\rho\; .
\ee
The coefficient $c$ is dimensionless 
and the Chern-Simons operator will therefore not decouple for large fermion
masses. A discussion that heavy fermion masses may not really decouple
from the low energy physics is given in \cite{banks}.
The Chern-Simons coefficient $c$ 
can be computed from the low energy
portion of the vacuum polarization graph in fig.4. One obtains
\be \label{eq4.17}
c = \frac{i}{3!}\epsilon_{\mu\nu\rho}\frac{\partial}{\partial q_\nu}
    \int \frac{d^3 p}{(2\pi)^3}\mbox{Tr}\left.\left[ S(p)\Lambda_\mu (p,p-q)
                                               S(p-q)\Lambda_\rho(p-q,p)\right]
                            \right|_{q=0}\; .
\ee
Here $S(p)$ is the free fermion lattice propagator which will
be specified later and $\Lambda_\mu$ is the photon vertex.
The integration in (\ref{eq4.17}) is understood to be taken over the
3-dimensional Brillouin zone.
A first observation is that due to gauge invariance the photon vertex may be
replaced in favour of the fermion propagator via the Ward identity
\be \label{eq4.17a}
\Lambda_\mu (p,p) = -i\frac{\partial}{\partial p_\mu} S^{-1} (p).
\ee
Upon differentiation with respect to $\partial/\partial q_\rho$, the
coefficient can be written as
\be \label{eq4.17b}
c = \frac{-i}{3!} \epsilon_{\mu\nu\rho}
                    \int \frac{d^3 p}{(2\pi)^3}\mbox{Tr}\left\{
    \left[S(p)\partial_\mu S^{-1}(p)\right]
\left[S(p)\partial_\nu S^{-1}(p)\right]
\left[S(p)\partial_\rho S^{-1}(p)\right]\right\}\; .
\ee
The free lattice propagators $S$ contains the lattice momenta $\sin(p)$.
Therefore the lattice integral (\ref{eq4.17b}) appears to be  
quite horrible to compute. 
By exploiting its topological properties its calculation will become 
tractable, however. The topological  significance of the above integral can
be seen by noting
that $S^{-1}$ may be generically written as
\bea \label{eq4.18}
S^{-1} & = & a(p) + i\vec{b}(p)\vec{\sigma} \nn \\ 
       & = &  N(p)\left[\cos (|\theta (p)|) +i \hat{\theta}\vec{\sigma} 
                        \sin (|\theta (p)|)\right] \nn \\
       & \equiv & N(p)V(p)
\eea 
where
\be \label{eq4.19}
           N(p) = \sqrt{a^2 + \vec{b}(p)\vec{b}(p)}\; , \; \;
\vec{\theta}(p) = \hat{b} \arctan (|\vec{b}|/a) \; .
\ee
Quantities with 
an arrow denote a 3-dimensional vector and with a hat the corresponding unit
vector. In this notation $V(p)$ is seen to be a $2\times 2$ unitary matrix.
The integral (\ref{eq4.17b}) does not depend on $N(p)$
provided that $S^{-1}(p)$ does not vanish. Thus $S$ and
$S^{-1}$ may be replaced everywhere by $V$ and $V^\dagger$. The matrix
$V$ describes a mapping from the torus $T^3$ to the sphere $S^3$. The integral
(\ref{eq4.17b}) is then nothing else but the winding number of this map.
Consequently it can only take integer values up to some normalization
constant. The 3-dimensional example we have worked out here can be
extended to arbitrary dimensions. 

To be specific we now take the usual fermion propagator for Wilson fermions
for a constant mass $m$,
\be \label{eq4.20}
S^{-1}(p) = \sum_{\mu=1}^3 i\sigma_\mu\sin p_\mu + m + r\sum_{\mu =1}^3\left(
            1-\cos p_\mu \right).
\ee

\begin{figure}                                                              
                                                                              
\vspace{7cm}
\includegraphics{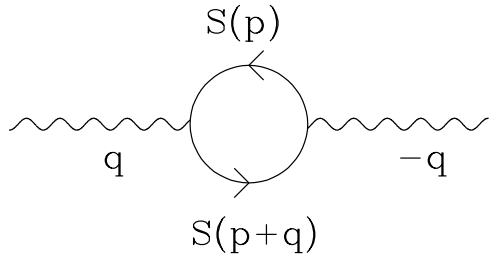}
\caption{ \label{fig4}                                                         
The vacuum polarization graph. The low energy portion of this graph leads to
the Chern-Simons coefficient to be calculated for the anomaly.
}                                                                              
\end{figure}                                                                   

The winding number can only change where $S^{-1} = 0$. This happens only
for momenta at the corners of the Brillouin zone and when the ratio
$m/r =0,2,4,6$ as can be seen from (\ref{eq2.8}). 
For the ratio $m/r\rightarrow \pm\infty$, $V(p)\rightarrow \pm 1$
and the integral vanishes. Therefore the Chern-Simons
coefficient will be zero for $m/r < 0$ and $m/r > 6$. Otherwise the
integral (\ref{eq4.17b}) is piecewise constant 
and gives only contributions for momenta
in an infinitesimal region around the Brillouin corners at the given 
values of $m/r$. To evaluate the integral we then only have to pick up
these contributions and sum over them. Since the change of the winding
number happens at the corners of the Brillouin zone we may expand the 
$\sin$ and $\cos$ function which leads us to evaluate
\bea \label{eq4.21}
\frac{dc}{dm} & = & -i\sum_{k=0}^3 (-1)^k \frac{d}{dm}
               \int \frac{d^3 p}{(2\pi)^3}\frac{m-2rk}
              {\left[ p^2 + (m-2rk)^2\right]^2} \nn \\
            & = & -\frac{i}{2\pi} \sum_{k=0}^3 (-1)^k \left( \begin{array}{c}
                     3 \\ k
                  \end{array} \right)\delta (m-2rk).
\eea
Our final task is then a trivial integration to get the coefficient as
\be \label{eq4.22}
c = -\frac{i}{4\pi} \sum_{k=0}^3 \left( \begin{array}{c}
                     3 \\ k
                  \end{array} \right)\frac{m-2rk}{|m-2rk|}.
\ee
We want to remark at this point that the above derivation can be extended
for arbitrary dimensions. The integral (\ref{eq4.17b}) describes then a map
from the torus $T^d$ onto the sphere $S^d$ and the homotopy classes
of these mappings are identified by integers. Therefore the whole
calculation of the Chern-Simons coefficient follows very closely the above
discussion.

We now want to apply the result found above for the domain wall model we
are interested in. 
We plot in fig.5 the integer part of the integral (\ref{eq4.17b})
as a function of $m/r$ (dotted line). The dashed line corresponds
to the chiral zeromode spectrum as discussed in section (2.1). A value 
of $+1$ means one chiral fermion with positive chirality, $-2$ means two
chiral fermions with negative chirality etc. Clearly the Chern-Simons
coefficient follows exactly the behaviour of the change of the chiral
zeromode spectrum with $m/r$. 

One peculiar feature of the Chern-Simons coefficient is that it is zero
for negative $m/r$. Since in the domain wall model $r$ is positive and
$m$ has different signs on the two sides of the wall, this means
that the Chern-Simons current flows only on one side of the wall.
This is to be contrasted with the continuum analysis which reveals that
the current flows with equal strength but opposite signs on the two
sides of the wall. Furthermore, if we take for example $m=r=1$ the value
of  the Chern-Simons coefficient is $c=-i/2\pi$ for $m>0$ which is exactly
twice the continuum value. Thus we find that the strength of the
Chern-Simons current depends on the regularization used as already
emphasized in \cite{colue}. Of course, the divergence of the Chern-Simons
current across the wall comes out the same in both calculations giving
the correct strength of the anomaly.
It is quite remarkable that the appearance of the anomaly on the lattice 
holds also for values of the domain wall mass at the order of the lattice
cut-off. The fact that the Chern-Simons current flows only on one
side of the wall motivates even more Shamir's approach to use free boundary
conditions with the signs of the Wilson coupling and the mass chosen such
that we will  have current flow.

The result for the 3-dimensional system discussed above can be generalized to
arbitrary dimensions. In $d=2n+1$ dimensions  the number of chiral
zeromodes bound to the domain wall is for  $2k<|\frac{m}{r}|<2k+2$, 
$0\le k\le d-1$
\be \label{eq4.23}
         \left( \begin{array}{c}
          d-1 \\
           k
         \end{array} \right)
\ee
with the chirality of the modes to be $(-1)^k\mbox{sign}(m)$.
The corresponding Chern-Simons current is
\be
\frac{J_\mu^{CS}(\mbox{lattice})}{J_\mu^{CS}(\mbox{continuum})}
= 2(-1)^k\left( \begin{array}{c}
          d-1 \\
           k
          \end{array} \right)
\ee
with the continuum Chern-Simons current in $d=2n+1$ dimensions given as
\be 
J_\mu^{CS}=\frac{i(-1)^n}{(4\pi)^nn!|m|}\epsilon_{\mu\alpha_1..\alpha_{2n+1}}
            F_{\alpha_1\alpha_2}...F_{\alpha_{2n}\alpha_{2n+1}}\; .
\ee

\begin{figure}                                                              
                                                                              
\vspace{7cm}
\includegraphics{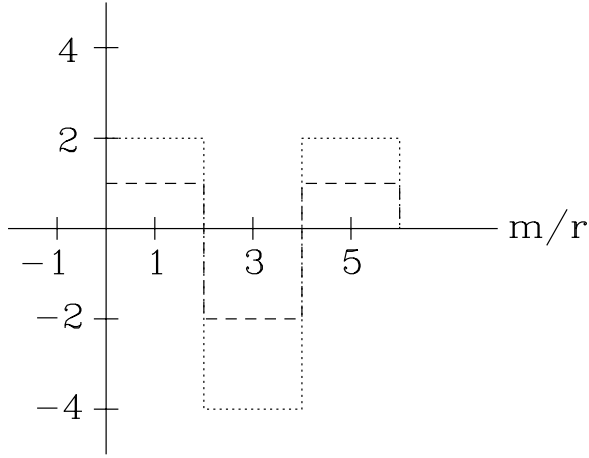}
\caption{ \label{fig5}                                                         
The integer part of the Chern-Simons coefficient (dotted line) 
as a function of $m/r$. The dashed
line indicates the chiral zeromode spectrum. $+1$ means 1 chiral zeromode
with positive chirality, $-2$ is two chiral zeromodes with negative chirality.
For $m/r < 0$ and $m/r > 2d$ the spectrum contains no chiral zeromode and
the Chern-Simons current vanishes, accordingly.
}                                                                              
\end{figure}                                                                   

The above scenario can be implemented on a finite lattice
of size $L^2L_s$.
The euclidean lattice action for free domain wall fermions is given by
\begin{eqnarray} \label{eq4.24}
S & = & \frac{1}{2}
     \sum_{z,\mu} \left[ \psibar_z \gamma_\mu \psi_{z+\mu}
                 -\psibar_{z+\mu}\gamma_\mu  \psi_z
\right] +m(s) \sum_z \psibar_z\psi_z \nonumber \\
 & + & \frac{r}{2}\sum_{z,\mu} \left[ -2\psibar_z\psi_z
+\psibar_z\psi_{z+\mu} + \psibar_{z+\mu} \psi_z 
               \right] 
\end{eqnarray}
with $z=(x,t,s)$ and $\mu=1,...,3$. 
We can couple external abelian gauge fields to the fermions
in close analogy to the Hamiltonian (\ref{eq3.3}) by making 
the finite lattice differences
gauge covariant, see eqs. (\ref{eq3.1},\ref{eq3.2}). 
Note that at this point the gauge fields are purely
external. How to write down a dynamical gauge field action will be discussed
in the next section. The lattice current is obtained by a gauge variation of the
action and reads \cite{kasmi}
\be \label{eq4.25}
j_z^\mu = \frac{1}{2}\left[\psibar_z \gamma_\mu U_{z,\mu} \psi_{z+\mu}
      + \psibar_{z+\mu}\gamma_\mu U_{z,\mu}^* \psi_z \right]+
  \frac{r}{2} \left[\psibar_zU_{z,\mu}\psi_{z+\mu} 
   - \psibar_{z+\mu}U_{z,\mu}^* \psi_z \right]\; .
\ee
We will choose a t-dependent external gauge field
\be \label{eq4.26}
U_{z,\mu=2} = \exp\{-iq\left[\frac{L}{2\pi}E_0\cos(\frac{2\pi}{L}(t-1))
  \right]\}
\ee
and make $E_0 \ll 1$ in order to stay in the low energy regime
and below the critical momentum $k_c$.
The $U$'s in the other directions have been set to one.

As we are considering free fermions in an external gauge field background,
the matrix elements of the 
current can be computed from the inverse fermion matrix using standard
numerical techniques like Conjugate Gradient for the matrix inversion.
Note, that in this way the computation of the current is 
--up to rounding errors--
exact. In particular, no simulation is involved.

Since the width of the wavefunction in the extra dimension is finite,
see fig.1, we will sum the current over the range in s corresponding to the
support $\Lambda_\psi$ of the wavefunction. Then we evaluate the divergence
\be \label{eq4.27}
<\partial_ij_i> \equiv \sum_{s\in \Lambda_\psi}
\partial_ij_i(t,x,s)\;\;, i=(t,x)\; .
\ee

According to the discussion for the Hamiltonian formalism we expect the
anomaly equation to be satisfied
\begin{equation} \label{eq4.28}
<\partial_ij_i> = \pm\frac{q^2}{2\pi}E_{eff}(t)
\end{equation}
\noindent where
the effective electric field $E_{eff}$ for small $E_0$ is given by
\begin{equation} \label{eq4.29}
E_{eff}=\frac{\sin(\frac{2\pi}{L})}{\frac{2\pi}{L}}E_0\sin(t-1)
\end{equation} 
and the sign is determined by the chirality of the mode. 
Choosing first the Wilson parameter $r=0$ and a charge
of $q=+1$, it is found that the divergence of
the current (\ref{eq4.27}) vanishes. This is perfectly consistent with the 
fact that for $r=0$ the doubler modes are still in the spectrum, cancelling
the anomaly.  

Turning the Wilson parameter on should change the picture. We expect the
doublers to become decoupled and the anomaly equation to be satisfied.
The divergences of the currents computed for charges of $q=3,4,5$ 
and chirality $+1,+1,-1$ are
shown in fig.6. Each of them follow the anomaly equation individually
up to a few $\%$.
Taking the sum of them the total divergence vanishes which corresponds to the
anomaly cancellation as predicted by the Pythagorean relation $3^2+4^2-5^2=0$.
In particular, the numerical results confirm that the Chern-Simons 
current flows only on side of the wall \cite{jansen2}.
Thus we see both, the individual anomaly for a fermion with a given charge
and the cancellation of the anomalies if the fermions are in an anomaly free
representation.

\begin{figure}                                                              
                                                                              
\vspace{9.2cm}
\includegraphics{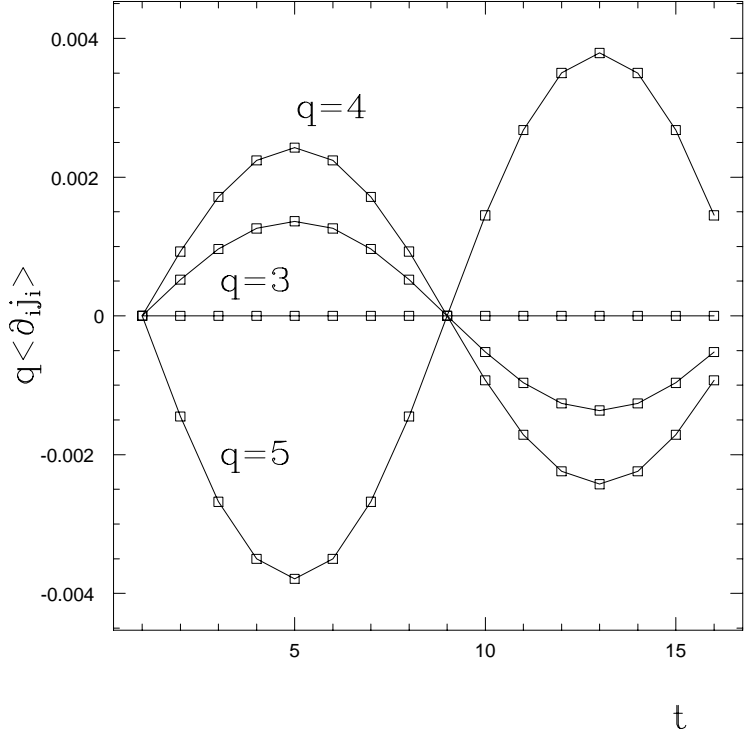}
\caption{ \label{fig6}                                                         
The divergence of the (1+1)-dimensional gauge current in an external gauge
field. The system consists of fermions with charge $q=3,4,5$ and chirality
$+,+,-$, respectively. Each of the currents obey the lattice anomaly
equation individually. The sum of the currents vanish due to the anomaly
cancellation.
}                                                                              
\end{figure}                                                                   

\section{Coupling to gauge fields}

Despite all the positive results we found in the previous sections the
crucial question remains, whether we can keep the chiral zeromodes
decoupled from each other
when the gauge fields are made dynamical. The danger is that the gauge fields
might induce an interaction between the two domain walls. 
We will discuss two proposals of coupling gauge fields to 
domain wall fermions. In a first attempt, the gauge field is
5-dimensional. The gauge couplings are chosen differently for the
4-dimensional fields and in the fifth direction. The hope in this approach is that by
tuning the 4-dimensional gauge coupling to zero and making the coupling
in the fifth direction strong at the same time, physics is confined to
4 dimensions. One then might perform the usual continuum limit in
4 dimensions while keeping the chiral structure of the theory.

We will see, however, that this path does not lead to the desired result,
namely a 4-dimensional chiral gauge theory. The failure of this approach 
suggests that one has to keep the gauge fields strictly 4-dimensional
from the very beginning.
In order not to couple both walls, the gauge fields
have to be switched off on one
of the domain walls such that only one wall is gauged. 
We will thus have a situation with identical 4-dimensional gauge fields
on a number of $s$-slices around a wall and $U=1$ in the complementary region
around the anti-wall. Clearly, at the boundary where both regions meet
we loose gauge invariance.
This can be repaired by
introducing a scalar (or St\"uckelberg) field at 
this boundary. However,
as we will see, 
there exist light mirror fermions living on the boundary. They can interact
with the chiral zeromode on the gauged domain wall 
thus rendering the theory vectorlike again. 

\subsection{Dimensional reduction}

In this 
first way to couple the gauge fields 
the gauge interaction is split into a purely 4-dimensional part and a piece
that contains the gauge fields in the fifth direction. Both parts are equipped 
with different gauge couplings.
For the possible phase structure and the question of whether one may
obtain a chiral gauge theory we will consider a 5-dimensional system. 
The reason
for making a departure from our 3-dimensional setup will become clear
later.
For the discussion 
it is sufficient in a first step to consider
the pure gauge theory alone. We will choose U(1)-gauge fields. 
The action becomes
\be \label{eq5.1}
   S(U) = \sum_{x,s}\left( \beta_4 \sum_{\mu,\nu=1}^4\mbox{Re} 
      (U^s_{x,\mu}U^s_{x+\mu,\nu}U^{*s}_{x+\nu,\mu}
       U^{*s}_{x,\nu})
 + \beta_{5}\sum_\mu \mbox{Re} (U^s_{x,\mu}
       V^s_{x+\mu}U^{*s+1}_{x,\mu}V^{*s}_x)
             \right).
\ee
Here $\beta_4 = 1/g_4^2$ is the 4-dimensional and
$\beta_{5} = 1/g_5^2$ the 5-dimensional inverse gauge coupling squared.
We have denoted with $U$ the gauge fields living in 4 dimensions and
with $V$ the ones in the extra dimension.
Thinking of $s$ more as a flavour space than an extra dimension, we have 
denoted the $s$-dependence with a superscript on the fields. Note that the
role of the $V$-fields look very much as that of scalar fields coupling
to 4-dimensional gauge fields in some peculiar flavour space. 
We are interested in 
the properties of this 5-dimensional gauge theory in the limit of large
$\beta_4$ corresponding to the continuum limit in 4 dimensions.

Letting the gauge fields be $U_{x,\mu}=e^{i\theta_{x,\mu}}$ and 
$V=e^{i\theta_{x,\mu}'}$ 
the path integral
is
\be \label{eq5.2}
Z = \int{\cal D}\theta_{x,\mu}{\cal D}\theta_{x,\mu}' 
       e^{-S(\theta_{x,\mu},\theta_{x,\mu}')}\; .
\ee
We consider this model in the meanfield approximation \cite{meanfield,funi}
which is obtained by inserting 
\be \label{eq5.3}
1 = \int{\cal D}v_{x,\mu}\int_{-i\infty}^{i\infty}{\cal D}\alpha \exp\left\{
                     \sum_{x,\mu}\left[
                     \alpha_{x,\mu} \left( v_{x,\mu} 
                   - \theta_{x,\mu}\right)\right]\right\}
\ee
in the path integral. In (\ref{eq5.3}) the mean field $v$ is a complex variable
living on the links of the lattice
and $\alpha$ an auxiliary field.
There is an analogous expression for $\theta'$. Now the integration
over the original link variables $\theta$ and $\theta'$ 
decouple and the path integral becomes
\be \label{eq5.4}
Z = \int{\cal D}v\int{\cal D}\alpha\exp\left\{\sum_{x,s}\left( 
                      \beta_4 \sum_{\mu,\nu=1}^4\mbox{Re}v_{x,\mu\nu} 
 + \beta_5\sum_\mu \mbox{Re} v_{x,\mu 5} 
             \right)+W(\alpha)+W(\alpha')-v\alpha-v'\alpha'\right\}
\ee
where
\be \label{eq5.6} 
W(\alpha) = \int_0^{2\pi} d\theta_{x,\mu} e^{i\alpha\theta_{x,\mu}}
\ee
and $v_{x,\mu\nu}$ denotes the product of the $v$'s around an elementary
plaquette in 4 dimensions and $v_{x,\mu 5}$ the corresponding expression in
the fifth direction.
A saddle point of the action in (\ref{eq5.4}) is obtained 
by solving
\be \label{eq5.7}
     \begin{array}{cccccc}
\frac{\partial S}{\partial v} & = & \alpha\; ; & 
\frac{\partial S}{\partial v'} & = & \alpha' \\
\frac{\partial W}{\partial \alpha} & = & v\; ; & 
\frac{\partial W}{\partial \alpha'} & = & v'\; . 
\end{array}
\ee
Searching for translational invariant solutions
by setting $v$ and $v'$ to constants,
the explicit equations at tree level are \cite{funi}
\bea \label{eq5.8}
\alpha  & = & 2(d-2)\beta_4 v^3 +2\beta_5 v'^2 v+2\beta_4 v \nn \\
\alpha' & = & \beta_5\left( 2(d-1)v^2 v' + 2v'\right)
\eea
and
\be \label{eq5.9}
     \begin{array}{cccccc}
 v & = & \frac{I_1 (\alpha)}{I_0 (\alpha)}\; ; &
 v' & = & \frac{I_1 (\alpha')}{I_0 (\alpha')}
\end{array}
\ee
where
\be \label{eq.5.10}
I_k (\alpha) = \frac{1}{\pi}\int_0^\pi d\theta \cos^k(\theta)
               e^{\alpha\cos(\theta)}\; .
\ee

\begin{figure}                                                              
                                                                              
\vspace{9.8cm}
\includegraphics{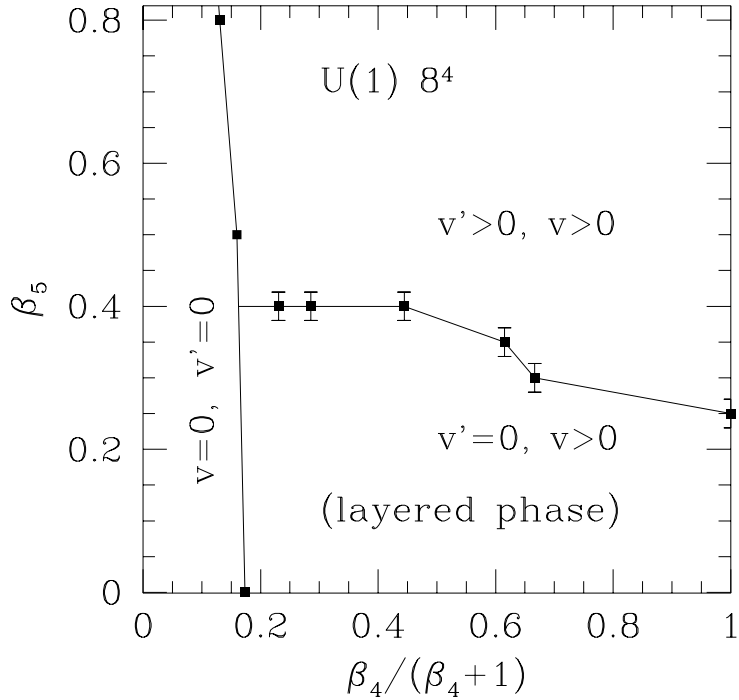}
\caption{ \label{fig7}                                                         
The phase diagram as obtained from a numerical simulation of the 
5-dimensional U(1) gauge theory for an $8^4$ lattice. Only the
portion for large $\beta_4$ is relevant for the continuum limit. 
}                                                                              
\end{figure}                                                                   

In principle  there  appear three different phases see \cite{funi} and
fig.7.
However, we are only interested in the large $\beta_4$ limit and will
not discuss the phase with $v=0$ and $v'=0$. For large $\beta_4$ one
finds a layered phase with $v>0$, $v'=0$ and separated from it 
by a phase transition at $\beta_5^c$ a symmetry
broken phase
with $v >0$
and $v' >0$. The layered phase is characterized by the fact
that a charged particle will only move along the layer and can not hop
between layers. Therefore if one would live inside a layer one would basically
only experience 4-dimensional physics.
In \cite{funi} it was argued that such
a phase only appears when the lower dimension is greater than two. Thus an
investigation of  
a 3-dimensional system would simply miss the layered phase
and might lead to wrong conclusions.
In fig.7
we show the phase diagram of the 5-dimensional pure U(1) gauge theory.
It is obtained by means of numerical simulations on an $8^4$ lattice.
As is predicted by the meanfield computation, there are indeed two
phases at large $\beta_4$. One of them has $v>0$ ans $v'>0$ and 
corresponds to the symmetry
broken phase. For $\beta_5 < \beta_5^c$ there appears the layered phase
which is the promising phase to obtain 4-dimensional physics. 
Taking the 1-loop corrections to the mean field equations into account,
the phase diagram remains stable.

The fermions are included via the action
\bea \label{eq5.11}
  S_{F} & = & \sum_{x,s} \sum_{\mu =1}^4
\psibar_{x}^s\left[(\partial_{x}(U)\gamma_\mu -\Delta_{x}(U) 
                          + m(s)\right]\psi_{x}^s \nn \\
    & + &  \frac{1}{2}\sum_{x,s}\psibar_{x}^s\left[
           V_{x}^s\delta_{s,s+\mu_5}(1+\gamma_5) 
          + V_{x}^{*s}\delta_{s-\mu_5,s}(1-\gamma_5) - 2\delta{s,s}
           \right] \psi_{x}^s
\eea
where we have separated the 4-dimensional part from the part in the fifth
direction and set $r=1$. 
Again we have written the $s$-label as a ``flavour'' label and $\mu_5$ denotes
a unit vector in the 5-direction.
In \cite{prades} it was shown that the fermions lead only
to a slight modification of the mean field equations and
that the phase structure depicted in
fig.7 remains stable when fermions are taken into account. 
The interesting question, of course, concerns the dynamics of the fermions
in the layered phase. 
One might get a hint by an inspection of the action itself
\cite{prd}. In the layered phase
$v'=0$ corresponding to set $V$ in the action (\ref{eq5.11}) 
to zero. But then the
action looks like a normal 4-dimensional {\em Wilson} action
in each layer. Indeed,
in \cite{prades} the fluctuations 
to the tree
level saddle point solution were computed to get the fermion propagator.
As a result 
one finds for every $s$-slice the 4-dimensional Wilson propagator. 
This means that the theory --though truly 4-dimensional--
is vectorlike in each s-layer. Therefore we have to throw away the region
$\beta_4 \gg 1$ and $\beta_5 < \beta_5^c$ as a possible 
corner of the phase diagram
for a construction of a chiral gauge theory from domain wall fermions.

In \cite{haoki} it could be proven by the hopping parameter expansion
that for $\beta_5 \ll 1$ the fermion propagator is parity invariant, with
the parity transformation defined such that for
$x_i\rightarrow -x_i$ with $x$ being the 4-dimensional coordinate
\be \label{eq5.12}
\psi\rightarrow \gamma_0\psi\; ,
\psibar\rightarrow\psibar\gamma_0\; .
\ee
Therefore there is a complete symmetry between left and right handed
modes. Thus the theory is completely vectorlike in each layer 
strengthening the above conclusions.

What remains is the possibility that $\beta_{5} > \beta_{5}^c$
\cite{prd}, i.e.
$v>0$, $v'>0$. In this situation we have symmetry breaking. In each
s slice there is a $G=U(1)\otimes U(1)$ symmetry such that the total symmetry
group is $G^{L_s}$
when the extent of the system in the fifth direction is $L_s$. 
For $v' > 0$ this symmetry is broken to its diagonal
subgroup $G$. Then we are left with $L_s-1$ massive gauge bosons with a mass
$m_G \propto v'$. However, there will remain one massless gaugefield
which does not depend on $s$.
This gauge field 
couples equally to the modes at the domain and the anti-domain wall.
Therefore,
although we still would have the zeromodes on the domain walls, they can
communicate via the massless gauge boson. We expect therefore 
the model again to be vectorlike.

\subsection{Waveguide model}

The previous section showed that starting with 5-dimensional gauge fields
does --most probably--
not lead to a chiral theory via dimensional reduction. Obviously
the gauge fields have to be strictly 4-dimensional from the very
beginning. This becomes a more natural point of view if one considers the extra 
dimension as a flavour space with a 
somewhat unusual flavour matrix. 
On the finite lattice with two domain walls not all s-slices (flavours) can
be gauged. This would immediately lead to the possibility that the zeromodes
on both domain walls communicate, rendering the theory vectorlike.

One might therefore try a scenario where only one of the
domain walls is gauged and can interact with the gauge fields. 
This corresponds to
taking a number of s-slices around a domain wall and put identical 
4-dimensional gauge fields
on each of these slices leaving the complementary $s$-slices ungauged. 
That the gauge fields have to be identical is 
again
motivated from the flavour space picture. In this way both walls are
completely shielded from each other. 
However, following the above prescription one notices that at the s-layer
where a gauged s-slice meets an ungauged layer gauge invariance is broken.

One possibility is that one does not worry about broken gauge invariance
\cite{parisi}. Indeed, several chiral fermion proposals exist that start
with broken gauge invariance \cite{rome}. 
It is then hoped that gauge invariance is restored in the continuum limit.
However,
we want to follow a path where gauge invariance is kept. Its 
restoration is actually very easy. One just has to remember how
the Yukawa-coupling in 
the Standard Model is made gauge invariant, namely by introducing scalar
(or St\"uckelberg) fields. 
From this point of view, the model with broken gauge invariance can be thought 
of as the gauge invariant model in the unitary gauge. Thus, both descriptions 
are actually completely equivalent \cite{jancapri}.
The proposal to couple gauge fields
in the domain wall model is
\begin{itemize}
\item keep the gauge fields strictly 4-dimensional, 
\item gauge only a number of s-slices around {\em one} domain wall,
\item introduce scalar fields at the boundary of the gauged region.
\end{itemize}
The gauge fields are then confined to a region in $s$ between the boundaries
where the scalar fields live. Thinking of electromagnetism this resembles
a situation where the electromagnetic waves are trapped in a waveguide
which suggests the name ``waveguide model'' that we will use from now on.

The introduction of the scalar fields gives rise to a coupling between
the fermions and these fields.
It is then natural to equip the resulting interaction with a 
Yukawa-coupling $y$.
To get a feeling for the physics of the model proposed,
one could first switch off the Yukawa 
coupling and set the gauge fields to one. In this situation the zeromode
spectrum can be obtained again from the Hamilton formalism. The result for
a 3-dimensional system is shown in fig.8a.
  
\begin{figure}                                                              
                                                                              
\vspace{13.0cm}
\includegraphics{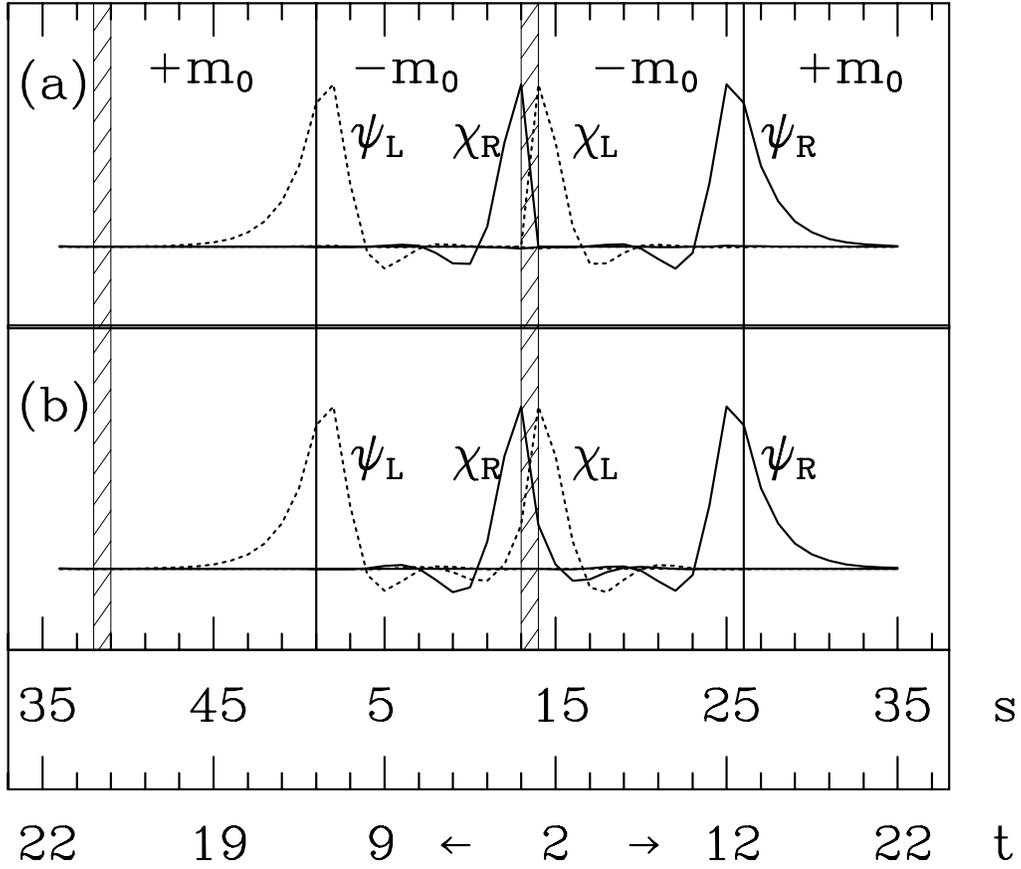}
\caption{ \label{fig8}                                                         
The zeromode spectrum in the waveguide model. The boundary where the scalar
fields live are indicated by the shaded stripes. The position of the
domain walls are given by the vertical bar. Fig.8a corresponds to $y=0$ and
fig.8b to $y>0$.
}                                                                              
\end{figure}                                                                   

The situation resembles a combination of the domain wall model and the
boundary fermion model. Setting $y=0$ 
cuts the waveguide region completely off and
produces open boundary conditions.
Due to our discussion in section (2.2) we expect to find chiral
zeromodes where the mass term is negative. 
This is indeed what happens. We find the original
chiral zeromode on the domain wall $\psi_L$ 
and a chiral surface mode --denoted in the following as ``mirror fermion''--
with opposite chirality $\chi_R$ at the boundary. 
A complementary picture is obtained in the s-region outside the waveguide.
   
Staying in a meanfield setup by giving the scalar field a vacuum expectation
value $v$, we show in fig.8b the spectrum for a non-zero value of $y$. Now the
mirror mode can leak out of the waveguide region and 
can combine with the mirror mode from outside the waveguide region to form
a Dirac particle. The mass of this particle is expected to follow the usual
perturbative relation $m_f=yv$. 

Let us make clear what we would like to achieve with the waveguide model.
Clearly one wants the chiral zeromode on the domain wall to survive. At the same
time, the mirror modes at the waveguide boundary should be heavy and therefore
decouple from the spectrum. This scenario looks hopeless if the mirror
fermion mass follows the perturbative relation. Approaching the phase transition
to the symmetric phase, $v$ (in lattice units) approaches zero. Then also the
mirror fermion becomes massless and can consequently not be decoupled.
Moreover, in the symmetric phase the mirror fermions are expected to be 
massless since there the vacuum expectation value vanishes identically. 
We would therefore end up with a variant of Montvay's
mirror fermion model \cite{montvay}.

What gives hope for the waveguide model is its resemblance to Yukawa-models
on the lattice. In fact, concentrating only on the waveguide boundary one
might consider the model as being just of the Yukawa-type. Such models have been
investigated intensively in the last years. The most surprising and
unexpected outcome of these investigations was the discovery of a new
phase at large, non-perturbative values of the 
Yukawa-coupling \cite{anna,su2pd}. 

When the Yukawa-coupling is very large, the fermions can easily combine with the
scalar fields and form massive bound states even in the symmetric
phase where the vacuum expectation value is zero. The masses of these states can
be at the order of the cutoff such that they decouple in the continuum
limit. What made the Yukawa models not successful for a regularization
of a chiral gauge theory was that the continuum limit in the strong coupling
region did not resemble the Standard Model at all. 
The fermion spectrum in the continuum limit consisted of only a neutral
Dirac fermion. The necessary charged fermion that would couple to the
left handed gauge field is simply missing and does not appear in the spectrum.
What is important is, that
on the other hand it was easy to give the neutral fermion a large mass.

For the domain wall model the situation is different in that we still would
have the chiral zeromode on the domain wall. Here the strong coupling phase 
would just serve to make the modes at the waveguide 
boundary heavy. The real physics
--and hopefully a chiral theory-- would appear along the domain wall. 
The main question in the domain wall approach to chiral fermions on the lattice
is then whether in the waveguide realization a strong coupling phase exists.

This makes it necessary to explore the phase diagram of the waveguide model.
In doing so we will impose some simplifications. First, we set the gauge fields
$U=1$. This is certainly justified as in the continuum limit the gauge coupling
becomes small and the gauge fields might be treated perturbatively.
Furthermore, the investigations of Yukawa models on the lattice
revealed the strong coupling region leaving out the gauge fields, too.
Therefore it appears to be sufficient to study the system with fermions and
scalar fields alone. Secondly, we will go back again to a 3-dimensional system
for the numerical simulations that will be presented
below. Experience from earlier work suggests that they resemble
4-dimensional systems quite closely \cite{jiri}.  In particular the
strong coupling behaviour could also be identified 
in this lower dimensional models. Thus the
appearance or non-appearance of a strong coupling region in the 3-dimensional
model would strongly point towards that such a region may or may not exist
also in higher dimensional models.

\begin{figure}                                                              
                                                                              
\vspace{10.0cm}
\includegraphics{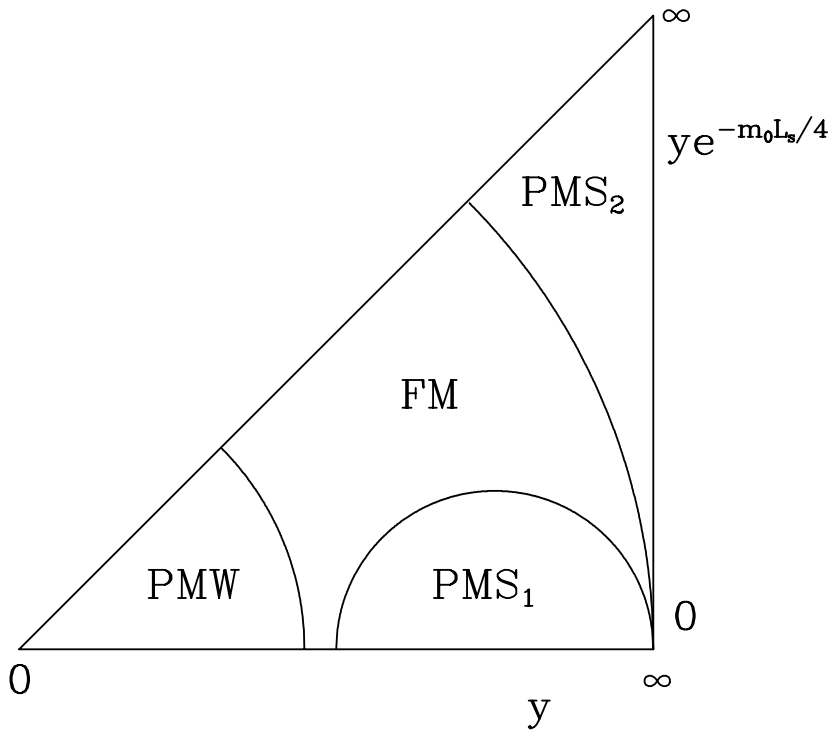}
\caption{ \label{fig9}                                                         
The phase diagram which would make the waveguide model successful.
}                                                                              
\end{figure}                                                                   

A sketch of the phase diagram which would make the waveguide model successful
is shown in fig.9.
There are two Yukawa-couplings. First there is the coupling we introduced
at the waveguide boundary to couple the scalar fields directly to the
fermions. Secondly, due to the exponential fall off, the domain wall
fermion will have an --exponentially suppressed-- overlap with the modes
at the waveguide boundary and hence couples to the scalar field with a
strength $y\exp\left(-m_0L_s/4\right)$. (The distance of the waveguide
boundary to the domain wall is $L_s/4$, see fig.8). The phase diagram in these
two couplings may have the following phases.
\begin{itemize}
\item {\bf FM} In the {\em ferromagnetic} phase the symmetry is spontaneously
broken and the scalar field assumes a vacuum expectation value $v>0$. 
If the model would be successful, there should be two regions
in the FM phase. One with weak coupling behaviour $m_F\approx yv$ and a strong
 coupling region with heavy fermions.
\item {\bf PMW} This is the {\em weak paramagnetic} phase. Here the symmetry
is restored and the vacuum expectation value $v=0$. In this phase the mirror
fermions living at the waveguide boundary are massless and do not decouple
from the spectrum. Therefore it is not possible to construct a chiral
gauge theory in this phase.
\item {\bf PMS$_1$} A {\em strong paramagnetic} phase. This is the phase we are 
looking for. Here due to the large Yukawa-coupling the binding of the 
fermions to the scalar fields at the waveguide boundary becomes so strong
that they combine and form massive bound states. 
At the same time the chiral zeromode
on the domain wall remains massless. Therefore the mirror fermions are
expected to decouple and we are --hopefully-- left with a chiral gauge theory
on the domain wall. 
\item {\bf PMS$_2$} Again a {\em strong paramagnetic} phase. However, this time
the extension of the system in the extra dimension is so short that the 
overlap of the chiral zeromode on the domain wall with the waveguide boundary
becomes non-negligible. Therefore the domain wall modes can also combine
with the scalar fields getting massive in this way. Clearly also in this
phase no chiral fermions can be obtained.
\end{itemize}
In the following, we will use PMW to indicate a weak coupling behaviour
of the fermion mass, $m_f\propto v$. PMS will indicate strong
coupling behaviour with massive bound states. Note, that in this way PMS and PMW
will not necessarily stand for weak and strong {\em values} of the
Yukawa-coupling $y$.
   
\subsection{The action of the waveguide model}

After the qualitative discussion of the previous subsection, we want to become
more concrete and will construct the lattice action suitable for
eventual simulations.
As we discussed above,
we take the same gauge field on all $s$-slices inside the waveguide. 
The index for the extra dimension will be chosen as a superscript to show
the resemblance to the flavour space picture.
We define gauge transformations on the fermion field as 
\be \label{eq5.13}
\begin{array}{rclrcll}

\Psi_x^s & \ra & g_x\Psi_x^s,&  \Psibar_x^s & \ra & \Psibar_x^s g_x^{\dagger} & \;\; 
               s \in WG , \nn \\
\Psi_x^s & \ra & \Psi_x^s,& \Psibar_x^s & \ra & \Psibar_x^s & \;\; 
                 s \not\in WG ,   \label{LIF}
\end{array}
\ee
\be  
           WG  =  \{s: s_0\leq s \leq s_0'\}     \label{WG}
\ee
with $g_x$ in a gauge group $G$. 
The detailed choice of the boundaries $s_0$
and $s_0'$ is not very important, provided they are sufficiently far
from the domain wall that the zeromode is exponentially small at
the waveguide boundary. For symmetry reasons we shall choose 
$s_0 = (L_s+2)/4+1$ and $s_0' = (3L_s+2)/4$, such that the right 
handed mode at $s=L_s/2+1$ is  located at the center of the
waveguide, see fig.\ 8.
With this choice we have to  take $L_s-2$ a multiple of four. 

Having made this division into a waveguide and its exterior, we note
that the model has a global $G\times G$ symmetry:
\be \label{eq5.14}
\begin{array}{rclrcll}
\Psi_x^s & \ra & g\Psi_x^s, & \;\Psibar_x^s& \ra & \Psibar_x^s g^{\dagger}, & \;\; 
                                  s \in WG,  \nn \\
\Psi_x^s & \ra & h\Psi_x^s,\; & \Psibar_x^s & \ra & \Psibar_x^s h^\dagger,& 
           \;\; s \not\in WG.
                                        \label{GLOBAL}
\end{array}
\ee
With our choice for
the position of the waveguide boundary, there is a symmetry
involving parity plus a reflection in the 
$s$-direction with respect to the plane $s=s_0-\frac{1}{2}=L_s/4+1$,
\be \label{eq.5.16}
\Psi_x^s \ra \gamma_d\Psi_{Px}^{L_s/2+2-s},   \label{PARITY}
\ee
with $Px = (-x_1,\cdots,-x_{d-1},x_d)$ the parity transform of $x$.

The layers, where gauge invariance is broken are now
$s_0-1$ to $s_0$ and from 
$s_0'$ to $s_0'+1$.
As discussed in the previous section, this will be repaired by the introduction
of scalar fields (or St\"uckelberg field) $V$ at the boundary of the waveguide.
Alternatively the scalar field $V$ might be interpreted as a component of the
gauge field in the extra dimension. In this sense the waveguide model is an
improved variation of the actions (\ref{eq5.1},\ref{eq5.11}) 
where the gauge fields have been made
5-dimensional. It corresponds to making the gauge coupling in the fifth
dimension $\beta_5 =\infty$ except at two s-slices, namely at $s=s_0$ and
at $s=s_0'$. We obtain the gauge invariant action (suppressing the
even dimensional index $x$)
\bea \label{eq5.17}
 S_\Psi 
       & = & \sum_{s\in WG} \Psibar^s(\dslash(U) -\Delta(U) + m^s)\Psi^s
 + \sum_{s\not\in WG}  \Psibar^s(\dslash-\Delta + m^s)\Psi^s \nn \\
  & - & \sum_{s\not=s_0-1,s_0'}[\Psibar^sP_L\Psi^{s+1}
                         + \Psibar^{s+1}P_R\Psi^s ]+\sum_s \Psibar^s\Psi^s
                    \label{SFUV}\\ 
 & - &  y(\Psibar^{s_0-1}VP_L\Psi^{s_0}  
   +  \Psibar^{s_0}V^{\dagger}P_R\Psi^{s_0-1})
   -   y(\Psibar^{s_0'}V^{\dagger}P_L\Psi^{s_0'+1} 
   + \Psibar^{s_0'+1}VP_R\Psi^{s_0'}),
   \nonumber
\eea
where we have supplied the Yukawa term  with a coupling constant $y$
as anticipated.
Note that we take the same scalar field at both waveguide boundaries.
Since we have chosen $r=1$ we have written projectors in the hopping terms 
in $s$, $P_{R(L)} = \frac{1}{2}(1 +(-)\gamma_5)$. 
$\dslash(U)=\gamma_\mu\partial_x(U)$ is the 
usual gauge covariant lattice Dirac operator , see (\ref{eq3.1},\ref{eq3.2}).
The field $V_x \in G$ is the scalar
field, which can be thought of as a (radially frozen) 
Higgs field, and which transforms as
\be \label{eq5.18}
         V_x \ra hV_xg_x^\dagger.   \label{LIB}
\ee
The transformation given in eq.~(\ref{PARITY}) remains a symmetry if 
$V$ transforms as
\be \label{eq5.19}
	 V_x \ra V^\dagger_{Px}.    \label{VPARITY}
\ee

Having introduced the scalar field, it is very natural to add a kinetic and
an interaction term for it. It is expected that through renormalization these
terms would be generated automatically. 
Since we have the scalar field to be radially
frozen, we choose the standard lattice form of the non-linear $\sigma$ model

\be \label{eq5.20}
S_V 
    = -\kappa \sum_{x,\mu} \mbox{Tr} 
       ( V_x U_{x,\mu} V_{x+\mu}^\dagger + h.c.).
\ee
The scalar hopping parameter $\kappa$ is proportional to the bare mass squared.
For the pure scalar theory we will have a broken phase for $\kappa > \kappa_c$
as well as a symmetric phase for $\kappa < \kappa_c$. 

\subsection{Mirror fermion representation of the model}
To get more insight into the mode spectrum of the model and to make the
interpretation of the surface modes $\chi_{L,R}$ in fig.8 as mirror
fermions more plausible, we 
rewrite the action as follows. Relabel the right and left 
handed fermion fields, $\Psi_{R,L}^s = P_{R,L}\Psi^s$
as 
\bea \label{eq5.21}
     \psi_{R}^t & = & \Psi_{R}^{s_0-1+t},  \;\; 
     \psi_{L}^t = \Psi_{L}^{s_0-t},  \nn \\
     \chi_{L}^t & = & \Psi_{L}^{s_0-1+t},  \;\; 
     \chi_{R}^t = \Psi_{R}^{s_0-t},
                         \label{NEWF}
\eea
and the same for  $\Psibar_{R,L} = \Psibar P_{L,R}$ (note the reversal of $L$
and $R$). The new label $t$ runs from $1$ to $L_t\equiv L_s/2$.
In  fig.\ 8 we 
have indicated this new labeling for the zeromode wave functions shown there.
With our choice 
for $s_0$, $s_0'$ and $L_s$ we can define a domain wall mass for both
fields $\psi$ and $\chi$, which is a step function in $t$ satisfying,
\be \label{eq5.22}
   \overline{\mu}^t = m^{s_0-1+t} = m^{s_0-t}.
\ee
With this relabeling the two domain wall zeromodes will reside
in the Dirac fermion field $\psi$, whereas the waveguide boundary zeromodes
will reside in $\chi$. After substituting eq. (\ref{NEWF}) into
eq. (\ref{SFUV}) with $U=1$, the action turns into 
\bea \label{eq5.23}
   S_{\psi\chi} 
       & = &
 \sum_{t=1}^{L_t} \left( \psibar^t \dslash \psi^t +   \chibar^t 
                  \dslash \chi^t +  
     \chibar^t( -\Delta + \overline{\mu}^t)\psi^t + 
     \psibar^t( -\Delta + \overline{\mu}^t)\chi^t 
                                                      \right)\nn \\
  & - & \sum_{t=1}^{L_t-1} \left(
     \psibar^t\chi^{t+1} + \chibar^{t+1}\psi^t \right) +
        \sum_t \left( \chibar^t\psi^t + \psibar^t\chi^t \right) \nn \\
  & - & y\chibar^1(VP_L + V^{\dagger}P_R)\chi^1
   -   y\psibar^{L_t}(V^{\dagger}P_L + VP_R)\psi^{L_t}.
                          \label{SMFV}
\eea

In this form, the action resembles that  of an $L_t$-flavour mirror
fermion model in the fashion of ref. \cite{montvay}, with $\psi$ the
fermion  and $\chi$ the mirror fermion field. In fact, for $L_s=2$ 
the hopping terms in $t$ are absent, $\overline{\mu}^t=0$ and
the model reduces to the mirror fermion model of ref. \cite{montvay}
with equal Yukawa  couplings for the fermion and the mirror fermion,
and a vanishing single-site mass term.
For $L_s>2$ the model has a more complicated mass matrix (i.e. non-diagonal
couplings among the flavours $s$ or $t$) and if the
model is going to be more successful  in decoupling the mirror fermion 
than the traditional mirror fermion approach, it must come from this
mass term.

The mass matrix for the $L_t$ flavours in the model is not diagonal
but this can be remedied by more rewriting. First we expand the fermion
fields in a plane wave basis, which diagonalizes the Dirac operator and
Wilson term, $\psi^s_x = \sum_p e^{ixp}\psi^s_p$, $\psibar^s_x = \sum_p
e^{-ixp}\psibar^s_p$. Here $\sum_p$ is a normalized sum over the 
momenta on the $d$ dimensional lattice, $\sum_p 1 = 1$. Then we can
write,  
\bea \label{eq5.24}
  S_{\psi\chi} & = &  \sum_{t=1}^{L_t} \sum_p\left( 
i\psibar^t_p \ssl_p \psi_p^t +  i \chibar_p^t \ssl_p \chi_p^t +  
     \chibar_p^t( w_p + \overline{\mu}^t)\psi_p^t 
  + \psibar_p^t( w_p + \overline{\mu}^t)\chi_p^t \right)\nn \\
   & - & \sum_{t=1}^{L_t-1} \left(
     \psibar_p^t\chi_p^{t+1} + \chibar_p^{t+1}\psi_p^t \right) 
  + \sum_t \left( \psibar_p^t\chi_p^t + \chibar_p^t\psi_p^t \right) \nn \\
  & - & y\sum_{pq}\left( \chibar_p^1(V_{p-q}P_L
        +  V^{\dagger}_{q-p}P_R)\chi_q^1 
   +   \psibar_p^{L_t}(V^{\dagger}_{q-p}P_L + V_{p-q}P_R)\psi_q^{L_t}
                  \right),
                          \label{SMPFV}
\eea
with $\ssl_p = \sum_\mu \gamma_\mu \sin(p_\mu)$,  $w_p$ the diagonal form
of the Wilson term, $w_p = \sum_\mu (1 - \cos(p_\mu))$  and $V_p$ the
Fourier transform of $V_x$. For $y=0$ the action has the schematic form 
\be \label{eq5.25}
   S_{\psi\chi}  =  ( \psibar \; \chibar)\left( \begin{array}{cc}
                           i\ssl   & M^\dagger \nn \\
                            M      & i\ssl  \end{array} \right)
                           \left( \begin{array}{c}
                             \psi \nn \\
                             \chi  \end{array} \right),
\ee
with $M$ a ($p$ dependent) matrix in flavour space, which can be read
off from eq. (\ref{SMPFV}).
This action can be diagonalized by making unitary transformations 
on $\psi$ and $\chi$,
\be \label{eq5.26}
\begin{array}{lcrlcr}
\omega^f & = & F^{\dagger}_{ft}\psi^t,\;\; &\omegabar^f & = &\psibar^tF_{tf},
  \nn \\
\xi^f & = & G^{\dagger}_{ft}\chi^t, \;\; &\xibar^f &= & \chibar^tG_{tf},

                  \label{MODE}
\end{array}
\ee
such that $G^{\dagger}_{fs} M_{st} F_{tg} = \mu_f\delta_{fg}$. 
The matrices $F$ and $G$ are eigenfunctions of $M^\dagger M$ and
$MM^\dagger$ respectively, labeled by the index $f$:
\be \label{eq5.27}
 (M^{\dagger} M)_{st} F_{tf} = |\mu_f|^2F_{sf},\hspace{1cm}  
 (M M^{\dagger})_{st} G_{tf} = |\mu_f|^2G_{sf}.
\ee
For suitable choices of the phases of the eigenfunctions, the $\mu_f$'s 
are  real.  Substituting the mode expansion (\ref{MODE}) into the 
action (\ref{SMPFV}) with the momentum label restored, we arrive at
\bea \label{eq5.28}
  && S_{\psi\chi} = \sum_{f=1}^{L_t} \sum_p\left( 
   \omegabar^f_p i\ssl_p \omega_p^f +   \xibar_p^f i\ssl_p \xi_p^f +  
     \xibar_p^f \mu_p^f\omega_p^f + \omegabar_p^f \mu_p^f \xi_p^f 
                          \label{SMODE}
                                                      \right)\\ 
  && -  y\sum_{fg,pq}\left[ 
  \xibar_p^f G^{p\dagger}_{f1}(V_{p-q}P_L + V^{\dagger}_{q-p}P_R)G^q_{1g}\xi_q^g 
   + \omegabar_p^f F^{p\dagger}_{f L_t}(V^{\dagger}_{q-p}P_L + V_{p-q}P_R)
                          F^q_{L_t g}\omega_q^g  \right].  \nonumber
\eea
In this representation of the model, it is seen that all fermion modes
$\omega^f$ and $\xi^f$
interact with the scalar field, but that their effective Yukawa-coupling
is determined by the magnitude of their wave function at the waveguide
boundaries $t=1$ and $t=L_t$. This leads then naturally to the introduction
of the two Yukawa-couplings we used in fig.9 to parametrize the phase diagram.
There we use the value of the wavefunction of the mirrormode at the waveguide
boundary which gives just $y$ and the value of the domain wall fermion at
the boundary which is $ye^{-m_0L_s/4}$.
For $y=0$ the model is seen to describe free, degenerate fermions and
mirror fermions with momentum dependent mass $\mu^f_p$ (for 
$\mu^f_p\not=0$, the eigenstates are $\omega^f_p + \xi^f_p$ and $\omega^f_p -
\xi^f_p$). Exactly one flavour, which we denote with $f=0$, has 
$\mu^0_p=0$ (up to terms exponentially suppressed in $L_s$)
for $|\hat{p}| < p_c$, where $p_c$ is the critical momentum 
defined in sect. 2.1.  For $r=1$ and $m_0$ close to 1, the
critical momentum is $p_c \approx \sqrt{2}$.
All other $\mu^f_p$ and also $\mu^0_p$ for
$p$ outside the critical momentum region, are $O(1)$ in lattice units.

This shows that for $y=0$  and momenta $|\hat{p}| \lsim  p_c$, 
the model contains 
a massless fermion, $\omega^0$, as well as a massless mirror fermion, $\xi^0$. 
All other modes ($f\not= 0$) as well as the species doublers have 
a mass of the order of the cutoff. 
The species doublers of the zeromode $f=0$ are massive because 
$\mu^0_p$ is $O(1)$ for momenta with $p_\mu= \pm \pi$. 

As was discussed already in section~5.2, fig.8 shows the 
$t$-dependence of the zeromodes $F^{t0}$ and $G^{t0}$ 
of the fermion (indicated by $\psi$ in the figure) and mirror fermion 
(indicated by $\chi$) for  the smallest momenta $|p| =\pi/L \ll p_c$. 
It shows that the zeromode for the
fermion is sharply peaked at $t=(L_t+1)/2$, i.e. at the domain
wall and the zeromode for the mirror fermion is localized at the boundary,
at $t=L_t$. The non-zeromodes, which are not shown in this figure, 
are not localized. 

\subsection{Large $y$ expansion}

The previous sections introduced the action of the waveguide model and through
a mode expansion exhibited its mode spectrum. 
In addition to the chiral zeromode on the domain wall, new mirror fermion modes
appeared at the wave guide boundary.
It can be expected that the mirror
zeromodes found for $y=0$, which are depicted in fig.8a, 
remain light and proportional to the vacuum expectation value
$v$ for small
but non-zero $y$. The interesting question is whether for large values of the
Yukawa-coupling the zeromodes at the waveguide boundary become massive and 
eventually decouple.

As fig.8 and the discussion of the previous section suggests, things may be
simplified by regarding a system with one waveguide only and to consider
the effective scalar-fermion model at the waveguide boundary alone. 
To get some insight into the spectrum at large $y$, one can try to perform a
$1/y$ expansion of the so truncated 
model \cite{gosha}. Again the gauge fields will be
set to $U=1$. The $1/y$ expansion is obtained by first relabeling the
fermion fields in $s$, while suppressing the even dimensional index
\be \label{eq5.29}
\begin{array}{cccccc}
\chi_s & = & \Psi_s\; , & s<s_0^{'},&   &  \\
\chi_{Rs_0^{'}} & = & \Psi_{Rs_0^{'}}\; ; &  \chi_{Ls_0^{'}}& = 
                     & \Psi_{Ls_0^{'}+1},\nn \\ 
\chi_s      & = & \Psi_{s+1}\; , & s>s_0^{'} & & \\
\psi_R      & = & \Psi_{R,s_0^{'}+1}\; ; & \psi_L     
            & = & \Psi_{L,s_0^{'}}\; .
\end{array}
\ee
By rescaling the field $\psi \rightarrow \frac{1}{\sqrt{y}}\psi$
the action becomes 
\bea \label{eq5.31}
S_F &=&\sum_s\left(\chibar_s\dslash\chi_s
   +\chibar_sa(s)(-\Delta+m-1)\chi_s\right)
 \nonumber \\
& + & \sum_s\left(\chibar_{Ls}\chi_{Rs+1}+\chibar_{Rs+1}\chi_{Ls}\right)
+\psibar_LV\psi_R+\psibar_RV^\dagger\psi_L \nonumber \\
& + & \sqrt{\alpha}\left[\psibar(-\Delta+m-1)\chi_{s_0^{'}}+
\chibar_{s_0^{'}}(-\Delta+m-1)\psi\right] 
+\alpha\psibar\dslash\psi,
\eea
where $\psi_{R,L} =P_{R,L}\psi\;$, $\chi_{R,L} =P_{R,L}\chi$,
\be \label{eq5.32}
\alpha = \frac{1}{y}\;\; \mbox{and}\;\; a(s)=1-\delta_{s,s_0^{'}}\; .
\ee
The sums over $s$ are to be taken in a region around $s_0^{'}$ 
neglecting all other defects
in the model which are thought of to be far away. 
In particular the domain wall mass is constant.

In this form it is seen that
for $y\rightarrow\infty$, the $\psi$ and the $\chi$ fields decouple. 
For the $\chi$-fields at the waveguide boundary one may 
derive the Dirac equation and look for solutions that simultaneously satisfy
the massless Dirac equation of the lower dimensional system.
Imposing plane waves in the even directions the analysis proceeds very
similar to the discussion in section~2.1. The Dirac equation reads in
momentum space
\be \label{eq5.33}
i\ssl(p)\chi_s+a(s)(m-1-F(p))\chi_s+P_R\chi_{s+1}+P_L\chi_{s-1} = 0.
\ee
Solving eq.(\ref{eq5.33}) for the $\chi$-fields,
one arrives
at normalizability conditions that dictate the chiral zeromode spectrum 
on the waveguide boundary \cite{gosha}. 
It turns out that in $4+1$ dimensions
one finds 7 righthanded and 8 lefthanded zeromodes \cite{gosha}
inside the waveguide, which are bound to the waveguide surface.
If the gauge fields are switched on,
this leads to a vectorlike spectrum if one takes the zeromode at the
domain wall into account.

An important question for a further exploration of the phase diagram
is whether the above result is stable against 
perturbations in $\alpha$. 
To answer this question we will now compute the
$\chi$-field propagator at the waveguide boundary to order $\alpha$. 
One may start by integrating out the
$\psi$-fields in (\ref{eq5.31}) which leads to an effective action

\bea \label{eq5.34}
S_\chi & = & \sum_s\left(\chibar_s\dslash\chi_s+
\chibar_sa(s)(-\Delta+m-1)\chi_s\right)
+\sum_s\left(\chibar_{s}P_R\chi_{s+1}+\chibar_{s+1}P_L\chi_{s}\right)
\nonumber \\
& - & \alpha\sum\chibar_{s_0^{'}}(-\Delta+m-1)\sum_{n=0}^\infty(-\alpha)^n
[(V^\dagger P_R+VP_L)\dslash ]^n(V^\dagger P_R+VP_L)(\Delta+m-1)\chi_{s_0^{'}}
 \nonumber \\
& = & \sum_{s,s'}\chibar_sS^{(0)-1}_{s,s'}\chi_{s'}
-\alpha\sum_x[\chibar_{s_0^{'}}(-\Delta+m-1)]_x(V^\dagger_xP_R+V_xP_L)
[(\Delta+m-1)\chi_{s_0^{'}}]_x \nonumber \\ 
&+& O(\alpha^2).
\eea
Here $S^{(0)}$ is the free $\chi$-field propagator which can be computed
exactly, see e.g. section~8.
To order $\alpha$, replacing $V$ by its expectation value $v{\bf 1}$, 
the Dirac equation reads now
\be \label{eq5.35}
i\ssl(p)\chi_s+a(s)(m-1-F(p))\chi_s+P_R\chi_{s+1}+P_L\chi_{s-1}
-\alpha v(m-1-F(p))^2\delta_{s,s_0^{'}}\chi_{s_0^{'}}=0.
\ee
For the symmetric phase with $v=0$ this is exactly the same equation as we
found for $y=\infty$. Therefore the result from the large-$y$ expansion is,
that in the symmetric phase we will find the same rich zeromode spectrum with
equal numbers of left and right handed fermions as we had at $y=\infty$. 
This is quite in contrast to the
anticipated phase diagram, fig.9, where we expected actually a 
symmetric phase with strong coupling behaviour.

For $v>0$ which corresponds to the broken phase
we have an extra piece in the Dirac equation at $s_0^{'}$,
$\alpha v(1-m+F(p))^2$.
This gives a mass to the mirror modes
of size $v/y$. 
We see that the weak coupling relation $m_f\propto yv$ is
replaced by $m_f\propto v/y$, i.e. $y\rightarrow 1/y$. 
In particular, the fermion masses are still proportional to $v$ and remain light
when 
approaching the phase transition, $v\rightarrow 0$. Passing the phase transition
we reach a symmetric phase with a vectorlike zeromode spectrum. 
In the discussion of the large-$y$ expansions we experienced a first surprise.
Since in the large-$y$ expansion all defects are thought to be very far
apart from each other, we effectively work in the limit $L_s \gg 1$. This
means that we are on the $ye^{-m_0L_s/4}\approx 0$-axis in the
phase diagram of fig.9. There we hoped to have the PMS$_1$ phase at 
$y\gg 1$. Instead we find a second symmetric phase with weak coupling behaviour
and an even richer --but vectorlike-- zeromode
spectrum than the one at weak Yukawa-couplings.

\subsection{Reduced model}

The mode expansion discussed in section~5.4 revealed more clearly than the
original action (\ref{eq5.17}) 
what the important modes of the waveguide model are.
There is, of course, the domain wall fermion living on the gauged wall. 
Furthermore, there are mirror fermion like states
that live on the waveguide boundary. These states can easily get light
and remain in the spectrum. We saw that this can happen for weak coupling as
well for very large coupling, which led to a revision of the phase diagram in
fig.9. 
However, at intermediate values of $y$, there might still be
still our candidate phase, PMS$_1$ where we can hope to construct a chiral
gauge theory on the domain wall at the end.
It became clear that it is very natural to think of the model as a mirror
fermion model along the lines of \cite{montvay} and that it would be the
peculiar
mass matrix which would be responsible for decoupling the mirror fermions
in contrast to the normal approach \cite{montvay}. 

As we will see below, numerical simulations in the waveguide model even
in the quenched approximation become very demanding. On the other hand
we could identify the important building blocks of the domain wall model,
the behaviour of which make the model succeed or let it fail, uniquely.
These building blocks are the
domain wall mode and the waveguide mode inside the waveguide. Furthermore,
these modes appear only in the low energy regime below the critical
momentum $p_c$. Above $p_c$ they become heavy and delocalize. 

One may therefore attempt to construct a reduced model
that shows the characteristic properties of the waveguide model
with these ingredients.
As we saw in section~5.5, the anticipated fermion spectrum for the reduced model
is not realized at very strong values of $y$. Therefore the reduced model 
only makes sense under the assumption that at intermediate $y$-values the
PMS$_1$ phase really exist.
We will use for this ``reduced'' model 
the domain wall zeromode $\omega^0$, 
the mirror mode $\xi^0$ at the
waveguide boundary, the scalar fields$V$ and the critical momentum $p_c$.
This leads to a model
\bea
   S^{red} & =  &
  \sum_{|{\hat p}|<p_c}[ i\omegabar^0_p \ssl_p \omega_p^0 + i\xibar_p^0 \ssl_p \xi_p^0 ]  
  +  y\sum_{|{\hat p}|,|{\hat q}|<p_c} \xibar_p^0(V^{\dagger}_{q-p}P_R 
+ V_{p-q}P_L)\xi_q^0.
                          \label{STOY}
\eea

In this reduced model the domain wall fermion $\omega^0$ is completely
decoupled from the mirror fermion $\xi^0$ and the scalar field $V$. The 
mirror mode is then coupled to the scalar field with a strength of the
Yukawa-coupling $y$. We now have exactly the situation which might lead
to a successful description of a chiral theory if the waveguide model
would work: An unaffected chiral zeromode on the domain wall and a Yukawa-like
theory on the waveguide boundary. The only question is, whether this
Yukawa-like theory possesses the strong symmetric phase or not. One may 
argue that the above model is an oversimplification of the full theory.
But one can give plausible arguments that this kind of reduction makes 
sense \cite{prd}. We will see further from the numerical simulations that
the reduced model behaves almost the same as the full model. 
This is so at least at the values of $y$ where the simulations discussed 
in the next section have been performed. We know, of course, from the
large-$y$ expansion, that the reduced model does not represent the
full model at $y\gg 1$.
Third,
if in the reduced model no PMS$_1$ phase exists, 
it will also not exist in the full
model. The only addition from the full model would be that 
the domain wall zeromode and the mirror mode start to interact. 
But then they either combine
to form a massive Dirac particle or the effective interaction is so weak
that we are back in the situation of the reduced model. The additional heavy
modes in the model, in particular the modes above the critical momentum $p_c$,
do not play an important role as their coupling can be expected to be
inversely proportional to their mass and hence is weak.

\section{Numerical simulations in the waveguide model}

Up to now we discussed the waveguide model at the two boundaries of the
phase diagram for small and large Yukawa-couplings. In both 
these limits it seems not to be possible to construct a chiral gauge
theory. This may, however, still be possible in the middle of the
phase diagram at intermediate values of the Yukawa-coupling. In this 
region both, the weak coupling as well as the strong coupling expansion
do not work. Therefore, in order to answer the question whether
the PMS$_1$ phase exists one has to rely on numerical simulations.

As already mentioned earlier, Yukawa models in 2 and 4 dimensions
have been subject to 
extensive investigations in the last few years. In these studies it became clear
that they could
either show a PMS$_1$ phase \cite{jiri} or not \cite{shrock},
depending on the choice of the Yukawa-coupling. 
Two examples of these kind of Yukawa models are the {\em local} coupling
\be \label{eq6.1}
  S_{lc} = y\sum_x\psibar_x(V_x P_R + V^*_x P_L) \psi_x ,
\ee
and the hypercubic coupling  
\be \label{eq6.2}
  S_{hc} = y\sum_{x}\frac{1}{4}\sum_b\psibar_x(V_{x-b} P_R 
                + V^*_{x-b} P_L) \psi_x
\ee
where we only show the Yukawa-part of the action.
In the case of the hypercubical coupling the
sum over $b$ means a sum over all points of a lattice hypercube
surrounding the point $x$. The model with
the local coupling is known to have a PMS phase whereas the one with 
hypercubical coupling does not. 

A form of the Yukawa-coupling
with a momentum cut-off as in the
waveguide model has so far not been investigated. Therefore the
question whether a PMS$_1$ phase exists or not remains undecided. 
Unfortunately the special form of the Yukawa-coupling as given in the 
wave guide model and in its reduced form prohibit an analytical 
treatment for intermediate values of $y$.  
Obviously a numerical study of the 5-dimensional system would be very
demanding. It turns out, however, that the strong coupling behaviour
we are after already appears in 2-dimensional Yukawa models. 
Even more, the phase diagrams of 2 and 4-dimensional models are very
similar so that it is a reasonable strategy to search for the PMS$_1$
phase in the waveguide model in 2 dimensions.

This path has been followed 
in \cite{prd}.
Most results are obtained in the quenched approximation, neglecting all
fermion loops. This seems to be sufficient for the question whether a PMS
phase exists. Although in the quenched approximation the 
phase transition lines shown in fig.9 do not really exist, the model should show
typical weak or strong coupling behaviour.
The experience from earlier studies of Yukawa models suggests two approaches
to identify or exclude a strong coupling symmetric phase. The first is the 
fermion mass spectrum and the second the eigenvalue spectrum. Both show
characteristic behaviour if a strong coupling phase exists.
The results of this numerical investigation 
will be discussed in the next two subsections.

The $(2+1)$-dimensional model is simulated using U(1) scalar fields. The pure
bosonic sector of the model consists therefore of a $XY$-model on the
boundary of the waveguide region.
One might worry that the $XY$-model we are going to look at is 
too different from the real 4-dimensional system 
we are interested in. There is a phase transition 
from a spin-wave to a Kosterlitz-Thouless phase at $\kappa\approx 0.5$
which is certainly in a different 
universality class compared to the phase transition in four
dimensions. In addition there is  no spontaneous symmetry breaking
in the $XY$-model. However, if on the finite lattice a vacuum expectation value
is defined through the standard rotation technique \cite{rotate},
\be \label{eq6.3}
v = <\frac{1}{L^2}|\sum_x V_x | >
\ee
then the
so defined vacuum expectation value $v$ is zero in the Kosterlitz-Thouless
phase but $v>0$ in the spinwave phase. In this way the 2-dimensional 
$XY$-model fakes the situation that appears in spontaneous symmetry
breaking in 4 dimensions. Increasing the lattice volume, $v$ approaches
zero also in the spinwave phase as it should. For a finite lattice we
expect therefore that the fermion mass follows the perturbative relation
$m_F \propto yv$, if the volume is not too large. We will in the following
refer to the spinwave phase as the broken and to the Kosterlitz-Thouless 
(or Vortex) phase as the symmetric phase.

All simulations presented below have been performed in a $(2+1)$-dimensional
setup with a lattice of size $L^2L_s$ with $L=12$, $L_s=26$ being typical
values of the lattice extensions. The Wilson parameter $r$ has always been set
to $r=1$ and the domain wall height was chosen to be $m_0=1$, too. 
Typically several thousand scalar field configurations have been obtained
using a cluster algorithm. The fermionic quantities have been obtained by
standard techniques like Conjugate Gradient or Lanczos. 
In order to avoid convergence problems of these algorithms antiperiodic
boundary conditions for one of the 2-dimensional indices have been chosen.

\subsection{Mass spectrum}

The fermion spectrum is a first indication for the presence of strong coupling
behaviour. Fermion masses can be obtained from the propagator 

\be \label{eq6.4}
  S^{st}(p) = L^{-2}\sum_{xy} e^{ip(x-y)}<\Psi^s_x\Psibar^t_y >,
\ee
which can be computed using the conjugate gradient algorithm
to find matrix elements of the inverse fermion matrix on a scalar field
background. Using the symmetry 
\be \label{eq6.5}
S_{RR}^{st}(p)=S_{LL}^{L_t+2-s,L_t+2-t}(Pp),
\ee
statistics can be enlarged by measuring the $LL$ and the $RR$ components
and averaging over them. Fermion masses are finally obtained by a fit to a free
fermion propagator ansatz
\be \label{eq6.6}
S(p)_{RR(LL)}=    
-iZ_F(\sin(p_1) -(+)i\sin(p_2))/(\sum_\mu \sin^2(p_\mu) + m_F^2).
\ee
The numerical results can be perfectly described by the ansatz (\ref{eq6.6})
(see fig.5 in \cite{prd}). 

\begin{figure}                                                              
                                                                              
\vspace{9.7cm}
\includegraphics{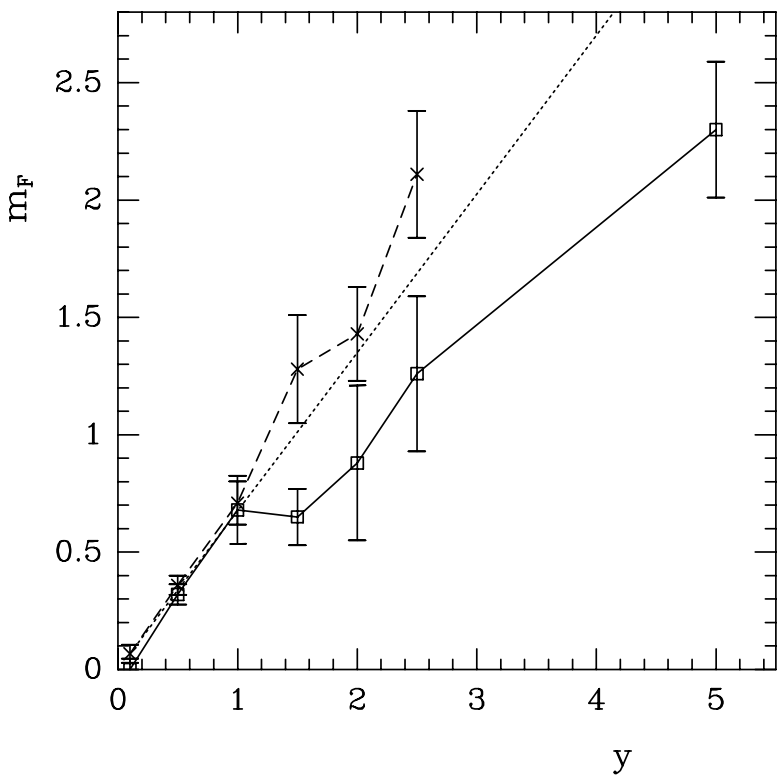}
\caption{ \label{fig10}                                                         
The fermion mass of the fermion at the waveguide boundary as a function
of the Yukawa-coupling y. Both, the fermion in the full model (boxes) and
the reduced model (crosses) are consistent with a linear behaviour, in
particular at small $y$. The dotted line is $yv$. All results are at
$\kappa=0.5$.
}                                                                              
\end{figure}                                                                   

What do we expect for the fermion masses having the phase diagram of fig.9
in mind?\\
{\em Weak coupling behaviour} For small values of the Yukawa-coupling
$m_f \propto yv$ with $v$ the vacuum expectation value of the scalar field
using (\ref{eq6.3}).
The proportionality of the fermion mass to $y$ 
could be confirmed in \cite{prd}, see fig.10. 
In the symmetric phase of the model the fermion mass turns out to be zero.
This excludes the small $y$ region for a 
construction of a chiral gauge theory as the mirror modes can
not be decoupled.

{\em strong coupling behaviour} For larger values of the Yukawa-coupling 
such that one does not enter the region where 
the large-$y$ expansion of section~5.5
is valid we
hope to find our favourite PMS$_1$ phase. From experience with models
showing the strong coupling phase, we expect the fermion mass to follow the
mean field strong coupling behaviour
\be \label{eq6.7}
m_F \propto z^{-2}\; ;\;z^2 = {\rm Re}\langle V_xV^{*}_{x+\mu}\rangle\; .
\ee

The quantity $z$ is non-zero in the symmetric as well as in the broken phase.
It is monotonically decreasing with decreasing $\kappa$. 
In the PMS$_1$ phase the fermion mass
measured from the numerical simulation would follow the behaviour in 
(\ref{eq6.7})
and would therefore {\em increase} with decreasing $\kappa$.
In fig.11 the behaviour of the fermion mass as
a function of $\kappa$ is shown at $y=2$ where one would expect the
strong coupling behaviour to show up if the model would  behave similar to
2-dimensional Yukawa models with local coupling.

\begin{figure}                                                              
                                                                              
\vspace{10.0cm}
\includegraphics{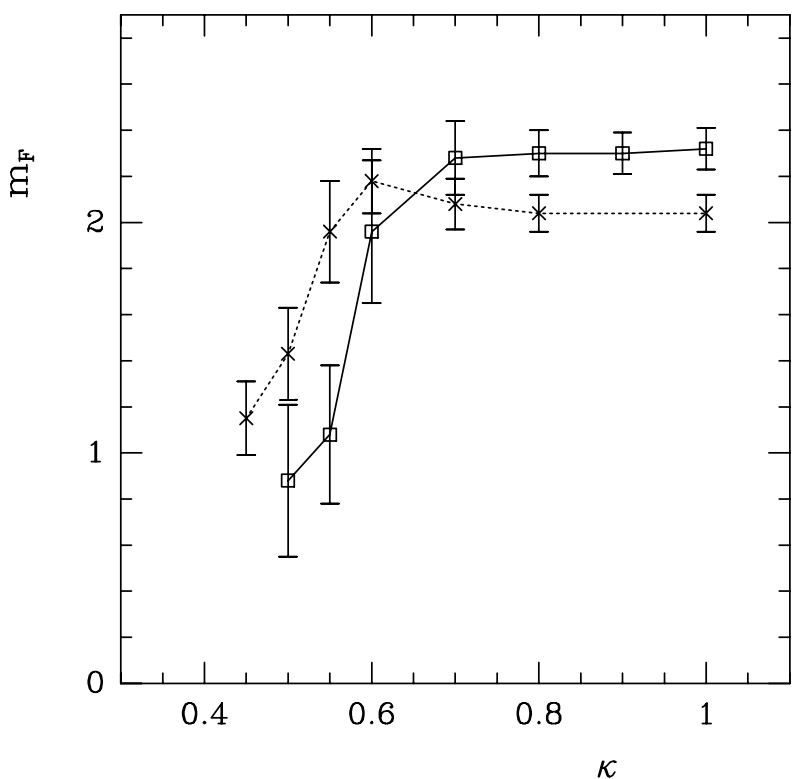}
\caption{ \label{fig11}                                                         
The fermion masses from numerical simulations at $y=2$ as function of $\kappa$.
The boxes correspond to the full and the crosses to the reduced model.
}                                                                              
\end{figure}                                                                   

In the figure, the boxes, connected by a dotted line to guide the eye,
correspond to the full waveguide model, the crosses to the reduced model.
The masses from both models are very similar. They stay first almost constant
at high values of $\kappa$ and then start to decrease when $\kappa$ is lowered.
This behaviour is in sharp contrast to the expected increase of the
masses if there would be a strong coupling region. Unfortunately, the numerical
simulations could not be driven into the symmetric phase 
($\kappa < \kappa_c =0.5$). The propagators
became too noisy and no reliable mass value could be obtained.

Nevertheless the picture is in clear contradiction with the scenario we
have hoped for. The mirror fermion masses become small approaching the
critical line. This is another indication that a PMS$_1$ phase 
does not exist in the waveguide model. 
Quite in contrary, the fact that fermions seem to become light in the
symmetric phase at $y=2$ point into the direction that the fermion spectrum
shows weak coupling behaviour.
One other important aspect of fig.10 and fig.11 is, that they show
the resemblance of 
the reduced model to the full model.
This similarity will be relied on when in
the next subsection the eigenvalue spectrum is determined,
which will give an additional hint that the PMS$_1$ phase in the
waveguide model is missing.

\subsection{Eigenvalue spectrum}

A complementary way to detect a strong coupling phase is the qualitative
behaviour of the eigenvalue spectrum for different Yukawa models
\cite{eigen}. The shape for a Yukawa-model with local (\ref{eq6.1}) 
and hypercubical (\ref{eq6.2})
couplings are quite characteristic. It would, of course be ideal to 
compute the eigenvalues in the full domain wall model. However, the matrix
from which the eigenvalues have to 
be determined is of size $2L_sL^2\otimes 2L_sL^2$.
To get all eigenvalues of this matrix is almost impossible for reasonably large
lattice sizes. We can, however, rely on the similarity of the reduced model
to the full waveguide model and try to determine the eigenvalues in the
reduced model at different values of $y$.

In fig.12 the eigenvalue spectra for the reduced domain wall model (a), the 
model with local (b) and with hypercubical (c) Yukawa-coupling is shown.
They have been obtained at $y=0.2, 1$ and $4$ from left to right
and were averaged over 5 scalar field configurations. The lattice size has
been $L=12$. For the waveguide model the eigenvalues at small
Yukawa-coupling are very similar for all three models and form a narrow
band. For $y=1$ there appear some differences in that for the waveguide and the
hypercubical model the band broadens whereas in the model with local
Yukawa-coupling they seem to develop a gap.

The difference becomes much more pronounced for $y=4$.  The local
Yukawa-coupling model shows a ring structure typical for a strong phase
\cite{eigen}. On the other hand
it is all too obvious that the domain wall model follows the trend of the
hypercubical model. Even at $y=4$ which is a quite strong Yukawa-coupling,
no sign of strong coupling behaviour can be seen. 
The eigenvalue distribution just broadens with increasing $y$.
The good representation of
the full model by the reduced one suggested in figs.10 and 11 leads to the 
conclusion that also in the full domain wall model the eigenvalues behave very 
similar. 

This is strengthened by the behaviour of the number of conjugate gradient
iterations to reach a given residue. This is determined by the condition
number of the fermion matrix. In the model with a local coupling which has the
strong symmetric phase, 
the number of conjugate gradient inversions first rises with
increasing $y$ but then, when entering the strong coupling region it
decreases again, due to the fact that the fermions become massive.
No such decrease can be seen in the full waveguide model. The number
of conjugate gradient inversions just continues to increase. This is, of course,
in full accordance with the results from the large-$y$ expansion and from the
fermion mass behaviour. They do not give a region 
with strong coupling behaviour but in 
contrast predict light fermions all the way to $y\rightarrow\infty$. 

\begin{figure}                                                              
\vspace{10.0cm}
\includegraphics{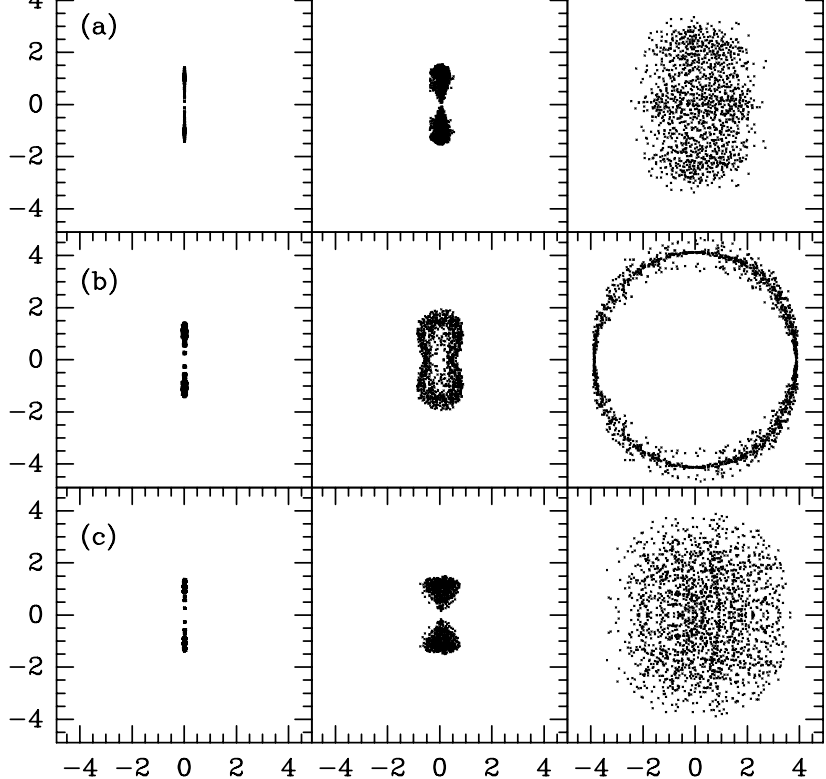}
                                                                             
\caption{ \label{fig12}                                                         
The eigenvalue spectrum for the reduced waveguide model (a), the Yukawa model
with local (b) and with hypercubical (c) Yukawa-couplings. The values of $y$
are $y=0.2,1.0$ and $4.0$ from the left to the right.
}                                                                              
\end{figure}                                                                   

\subsection{Dynamical fermions}

All the numerical results discussed in the previous section were obtained
in the quenched approximation. In the full unquenched model there
would have been a single right handed fermion coupled to the U(1) gauge field
if the fermions would have been in an anomalous representation.
However, the model in the
quenched approximation does not distinguish between an anomalous and an
anomaly free representation. We could have equally well added an extra
fermion species $S=\chibar M^\dagger \chi$ to the original action
$S=\psibar M\psi$ with $\psi$ and $\chi$ having opposite handedness. 
In the quenched approximation the extra fermion $\chi$ is irrelevant and
the quenched approximation of 
the full model with $\psi$ and $\chi$ reduces to the one studied
in the previous sections. 

Of course, it would have been desirable to study the waveguide model
also dynamically. This was also attempted. 
A Standard Hybrid Monte Carlo algorithm \cite{hmc} was used in the simulations.
Using the scalar vacuum expectation value $v$ as an order parameter, the
unquenched simulations revealed the 
weak symmetric phase for small values of $y$, where $v=0$. 
Keeping $\kappa$ fixed, the phase diagram in fig.9 suggests that by increasing
$y$ one would enter a symmetry broken phase with $v>0$. This pictures is
motivated by the ordering effect of the fermions which shift the phase
transition of the pure bosonic theory towards smaller values of $\kappa$.
Increasing $y$ even further, $v$ becomes zero 
for some critical Yukawa-coupling $y_c$,
if one enters a second symmetric phase.
Whether this is a phase with strong coupling behaviour could then 
be clearified 
by computing the fermion spectrum.

This scenario can be tested by numerically computing $v$
using the rotation technique. 
Unfortunately, for medium size values of $y\approx 1$,  the step size
in the Hybrid Monte Carlo algorithm had to be decreased by an order
of magnitude resulting in large autocorrelation times. This became even
worse when increasing $y$ further such that no reliable simulation
could be performed in the interesting $y$ region. The bad behaviour of
the Hybrid Monte Carlo algorithm is, however, consistent with the development
of the eigenvalue spectrum discussed in the previous section.  
If a strong symmetric phase exists one would expect the behaviour
of the Hybrid Monte Carlo algorithm to become much better again since
the fermion would have become massive. It might be an interesting
prospect to attack the difficult region in $y$ again by using the
recently suggested new simulation algorithm for fermions by L\"uscher
\cite{boson}.

\section{Infinite extra dimension}

As we have seen in the previous sections, our attempts to achieve a chiral
gauge theory with domain wall fermions seem to fail. The effect of the
anti-domain wall can not be switched off in case of 5-dimensional gauge
fields. In the waveguide model where gauge fields are kept 4-dimensional the
anti-wall can be shielded from the wall but only for the expense of 
introducing a wave guide wall. New light mirror fermion
modes appear on this wall
rendering the theory vectorlike again.

We found on the other hand in section~2.1 that on the infinite lattice
there is only one chiral zeromode and no necessity for introducing a
second wall emerges. One might first discard an infinite lattice as not leading
to a practical proposal for simulations. This is certainly the case for
the euclidean action formulation of a theory. Narayanan and Neuberger 
\cite{nane,nara} 
suggested to take a different route of thinking.
They regard the extra dimension as the ``time-axis'' of a $(4+1)$-dimensional
system and propose to study the 4-dimensional Hamiltonian of this theory.
The extra dimension is thought of being kept strictly infinite and as
such it fulfills the sole
purpose to project onto the ground state of the system.
There will be two Hamiltonians $H^\pm$ corresponding to the two sides of the
wall and accordingly there will also be
two ground states $|0\pm>$. The theory can then be defined as the overlap
of these two ground states.

Also in this approach the gauge fields are kept 4-dimensional. As
opposed to the waveguide model there is no need of adding an extra wall
and thus no additional fields have to be introduced. 
To keep the extra dimension infinite may also be motivated from the
index of the fermion matrix $M$. What one would like to achieve in order to
obtain a chiral theory is that the fermion matrices $M$ and $M^\dagger$,
belonging to the left and righthanded modes, respectively, 
have different indices. In particular, $M$ should have index 1 and
$M^\dagger$ index 0, which would correspond to a single chiral fermion
as an eigenstate of $M$. Such a situation can only be reached when the matrices
$M$ and $M^\dagger$ are infinite. Of course, in the domain wall model 
with a finite $M$ it was hoped that   
one can keep both chiral modes on the lattice such that they can be 
separated and shielded from each other. 
The model proposed in \cite{nane} can be considered
as a combination of domain wall fermions as suggested by Kaplan \cite{david}
and the idea of infinitely many regulator fields advocated by Frolov and 
Slavnov \cite{frolov}. There it was suggested to introduce infinitely
many Pauli-Villars fermion fields to regulate a chiral gauge theory.
Interpreting the extra dimension as a flavour space, the different flavours
can be associated with new fields that have been introduced. Choosing the
extra dimension to be infinite, these new fields play the role of the
infinitely many Pauli-Villars fields in the approach of \cite{frolov}.

To be specific, we will consider the following 4-dimensional action 
\bea \label{eq7.1}
  S_{F} & = & \sum_{x,s} \sum_{\mu =1}^4
\psibar_{x}^s\left[(\dslash(U)_x -\Delta_{x}(U) 
                          + m(s)\right]\psi_{x}^s \nonumber \\
    & + &  \frac{1}{2}\sum_{x,s}\psibar_{x}^s\left[
           \delta_{s,s+\mu_5}(1+\gamma_5) 
          + \delta_{s-\mu_5,s}(1-\gamma_5) - 2\delta{s,s}
           \right] \psi_{x}^s\; .
\eea
This action corresponds to the action (\ref{eq5.11}) 
with the 5-dimensional gauge fields
$V$ set to unity and the Wilson coupling $r$ set to $r=1$. 
The domain wall mass is chosen as
\be \label{eq7.2}
m(s) = d - m\mbox{sign}(s+\frac{1}{2})
\ee
with $d$ the (odd) dimension of the system.
Note that
the wall in this setup has been chosen to be in the middle of the link
connecting the regions in $s$ with positive and negative domain wall mass.
Of course,
keeping the summation over $s$ infinite leads to divergences. However,
these divergences come from the bulk and one may define an interface 
effective action
\be \label{eq7.3}
S_i(U) = S_{eff}(U) -\frac{1}{2}
        \left[S_{eff}^{+}(U)+S_{eff}^{-}(U)\right].
\ee
$S_{eff}^\pm (U)$ are the bulk actions for a constant domain wall mass
to the right and the left of the location of the domain wall. 
The interface effective action is defined on the wall. It is 
now finite and can be implemented 
order by order in perturbation theory. 

Staying in the euclidean action formulation would, however, not lead to
any practical way for numerical simulations. 
The path Narayanan and Neuberger are following is to construct
$S_i(U)$ from a vacuum overlap formula. This is obtained in a first step by
constructing the transfer matrix $\hat{T}=e^{\hat{H}}$ 
(with the lattice spacing $a=1$) from the lattice Hamiltonian $\hat{H}$,
$\hat{H} = \hat{a}^\dagger H\hat{a}$ with $\hat{a}^\dagger$ and $\hat{a}$ are
the usual fermionic creation and annihilation operators 
and $H$ is represented by
a matrix, the explicit form of which will be given below.

For Wilson fermions with $r=1$ the transfer matrix
describes the evolution of the system from one timeslice to the next 
and therefore over the smallest distance on the lattice.
As the transfer matrix contains the full physical information of a system,
a description of a model in terms of the transfer matrix is completely
equivalent to its path integral formulation. 
Indeed, the path integral $Z$ may
be written with the usual trace formula $Z=Tr\hat{T}^N$ when $N$ is the extension
of the lattice in the ``time'' direction.   
The construction of the transfer matrix, which is a non-trivial task,
was performed by Narayanan and Neuberger, following 
\cite{luscherham} closely. They find
the transfer matrices on both sides of the wall, $\hat{T}_{+}$
and $\hat{T}_{-}$. $\hat{T}_{+}$ and $\hat{T}_{-}$
are the main building blocks of the theory in the
infinite extra dimension formulation 
and for the construction of the overlap formula.

As a result a representation of
the interface effective action in terms of the transfer matrices can be given
\be \label{eq7.4}
e^{S_i(U)} = \lim_{s\rightarrow\infty} \frac{
           <b-|(\hat{T}_{-})^{s-1}(\hat{T}_{+})^{s-1}|b+ > }{
\sqrt{<b-|(\hat{T}_{-})^{2s-2}|b->
      <b+|(\hat{T}_{+})^{2s-2}|b+ >} 
}
\ee
where $<b\pm |$ denote some boundary conditions at $s=\pm\infty$. 
Choosing these boundary conditions as 
\be \label{eq7.5}
|b\pm > = |0\pm >
\ee
in the $s\rightarrow\infty$ limit the ground state is projected out in
(\ref{eq7.4}) and the interface effective action is given by a simple
overlap formula
\be \label{eq7.6}
e^{S_i(U)} = < 0- | 0+ >\; .
\ee

To get a feeling for the formula (\ref{eq7.6}) which corresponds to the
--infinite-- lattice effective action, we will now as a first step discuss
its continuum limit for a $(2+1)$-dimensional system. 
We will check in continuum perturbation theory
whether we will find the anomaly we expect, if the overlap formula 
correctly describes the chiral gauge theory it was constructed for. 
The Hamiltonians for the left and the right sides of the wall,
$H_{-}$ and $H_{+}$, respectively, are in momentum space 
\be \label{eq7.7}
-H_\pm(p,q) = H_0^\pm(p)\delta_{p,q} + \int\frac{d^2k}{(2\pi )^2}\left[
                 z_k\tau\delta_{q,p+k}+z_k^{*}\tau^\dagger\delta_{q,p+k}\right]
\ee
where the unperturbed Hamiltonians $H_0^\pm$ are diagonal in momentum space
\be \label{eq7.8}
H_0^\pm(p) = \left( \begin{array}{cc}
                     \mp m & p_1-ip_2 \\
                     p_1+ip_2 & \pm m
                     \end{array} \right)\; ,
\ee
\be \label{eq7.9}
\tau = \left( \begin{array}{cc}
                     0 & 0 \\
                     1 & 0
                     \end{array} \right)
\ee
and 
\be \label{eq7.10} 
z_k = \int d^2xe^{ikx}\left[A_1(x) + iA_2(x)\right]\; .
\ee
If we denote by $\Psi_p^\pm$ the set of eigenvectors belonging to positive
eigenvalues of the full Hamiltonian
\be \label{7.11}
-H_\pm\Psi^\pm_p = \Lambda^\pm_p\Psi^\pm_p\; ;\;\; \Lambda^\pm_p > 0
\ee
we can define an overlap matrix as
\be \label{eq7.12}
O_{pq} = \Psi_p^{-\dagger}\Psi_q^{+}\; .
\ee
The continuum interface effective action is then given by
\be \label{eq7.13}
e^{S_i(A)} = \mbox{det}O\; .
\ee
To proceed we will now sketch the perturbative evaluation of $S_i(A)$.
For this purpose we expand the fields
\bea \label{eq7.14}
\Psi^{+}_p & = & \psi_p^{+} + \psi_q^{+}A_{qp}  + \chi_q^{+}B_{qp} \nn \\
\Psi^{-}_p & = & \psi_p^{-} + \psi_q^{-}C_{qp}  + \chi_q^{-}D_{qp} 
\eea
where $\psi^\pm$ and $\chi^\pm$ are the eigenvectors corresponding to the
positive and negative eigenvalues of the unperturbed Hamiltonian $H_0^\pm$,
respectively. $A$, $B$, $C$ and $D$ are coefficients which are linear in
the gauge fields. The eigenvalues are also expanded
\be \label{eq7.15}
\Lambda_p^\pm = \lambda_p + \lambda^{\pm (1)}_p
\ee
with $\lambda_p$ the eigenvalues belonging to the free Hamiltonian. 
The perturbation Hamiltonian can be read off from (\ref{eq7.7})
\be \label{eq7.16}
-H_1(p,q) = \left[ z_k\tau\delta_{q,p+k}
                  +z_k^{*}\tau^\dagger\delta_{q,p+k}\right] 
\ee
and the coefficients and the eigenvalues can be obtained from standard
Schr\"odinger perturbation theory
\bea \label{eq7.17}
\lambda_p^{\pm (1)} & = & \psi^{\pm\dagger}_pH_1\psi^\pm_p \nn \\
A_{pq}              & = & \frac{\psi^{+\dagger}_pH_1\psi^{+}_q}
                               {\lambda_q - \lambda_p}\;\;p\ne q \nn \\
B_{pq}              & = & \frac{\chi^{+\dagger}_pH_1\psi^+_q}
                               {\lambda_q + \lambda_p} \nn \\
C_{pq}              & = & \frac{\psi^{-\dagger}_pH_1\psi^-_q}
                               {\lambda_q - \lambda_p}\;\;p\ne q \nn \\
D_{pq}              & = & \frac{\chi^{-\dagger}_pH_1\psi^-_q}
                               {\lambda_q + \lambda_p} \; .
\eea
The diagonal terms $A_{pp}$ and $C_{pp}$ in (\ref{eq7.17})
are undetermined. This can be partly
repaired by imposing orthonormality of the eigenvectors which leads to
\be \label{eq7.18}
A_{pp} + A^{*}_{pp} =0\; ;\;\; C_{pp} + C^{*}_{pp}=0
\ee
which fixes the real part of both coefficients. 
However, the imaginary part of $A$ and $C$ remains undetermined
and therefore there is a phase ambiguity in the interface effective action.
This phase ambiguity reflects the fact that it is the imaginary 
part of the complex action where the chiral nature is hidden \cite{alva}.
The real part of the action is vectorlike. This can also be seen from 
(\ref{eq7.4}).
Taking there periodic boundary conditions in the extra $s$ direction,
in the limit $s\rightarrow\infty$
\be \label{eq7.19}
e^{S_i(U)} = |<0-|0+>|^2\; .
\ee
The effective action (\ref{eq7.19}) is real and corresponds to the
domain wall setup with two chiral zeromodes of opposite chirality. Thus we see
that the imaginary part of the action is the crucial ingredient for 
chiral gauge theories.
One proposal to compute the imaginary part of the action can be found in
\cite{schierholz}.

The still undetermined phases for the interface effective action
have to be fixed. 
This can be done in perturbation theory by
demanding that the projection of the true eigenstate onto the
perturbed one $\Psi_p^{\pm^\dagger}\psi^\pm_p$ being real. This amounts
to Brillouin-Wigner perturbation theory. Of course, also for 
a non-perturbative investigation
of a chiral gauge theory the phase ambiguity appears. The authors
in \cite{nane} propose to take the same phase fixing condition 
as in perturbation
theory. The advantage of this condition is that it can be easily implemented
and thus the theory is completely determined. It is, however, an important
--and still open-- question whether the Brillouin-Wigner condition is suitable
also for non-perturbative and strongly fluctuating gauge field configurations.

To arrive eventually at the anomaly one has 
to go to second order in perturbation
theory. This can be done \cite{nane} and as a result the anomaly is given as
\be \label{eq7.20} 
\sum_\mu\partial_\mu\frac{\delta I}{\delta A_\mu(x)}
\ee
where
\be \label{eq7.21}
I =\frac{1}{16\pi}\sum_p\left[\frac{p_1+ip_2}{p_1-ip_2}z_p^{*}z_{-p}^{*}
                             -\frac{p_1-ip_2}{p_1+ip_2}z_pz_{-p}\right]
\ee
and $z$ is given in (\ref{eq7.10}),
\be \label{eq7.22}
z_p = A_{1,p} +iA_{2,p}\; ; z_{-p}^{*} = A_{1,p} -iA_{2,p}\; .
\ee
Thus the anomaly becomes
\be \label{eq7.23}
\frac{1}{4\pi}\left(p_2A_{1,p}-p_1A_{2,p}\right) = 
        -\frac{1}{4\pi}F_{12}(p)\; .
\ee
Eq.(\ref{eq7.23}) is the continuum form for the anomaly on the wall from the
overlap formula (\ref{eq7.6}). Since in deriving this equation the variation of
the action with respect to the gauge fields was taken for the effective
interface action we should have expected to find the anomaly which
corresponds to the chiral mode on the wall. This would be the consistent anomaly
as we discussed in section~4. In fact, the strength of the anomaly in 
(\ref{eq7.23}) is
$1/4\pi$ which is just half of the value of the covariant anomaly
we have discussed earlier for the domain wall system. Thus it corresponds
to the consistent anomaly which is the correct result.

As a next step it would be certainly interesting to see, whether the
overlap formalism can be taken over to the lattice.
The transfermatrix can be explicitly constructed
\be \label{eq7.24}
e^{H_\pm} = \left(
           \begin{array}{cc}
           \frac{1}{B^\pm} & \frac{1}{B^\pm}C \\
           C^\dagger\frac{1}{B^\pm} & C^\dagger\frac{1}{B^\pm}C + B^\pm
           \end{array}   \right)\; .
\ee
The matrices $B$ and $C$ depend on the gauge fields and their explicit forms
are
\be \label{eq7.25}
B^\pm = (d+1\mp m) -\frac{1}{2} \sum_\mu\left
       [\delta_{x,x+\mu}U_{x,\mu}+\delta_{x,x-\mu}U^*_{x-\mu,\mu}\right]
\ee
\be \label{eq7.26}
C = \frac{1}{2}\sum_\mu\left[\delta_{x,x+\mu}U_{x,\mu} - \delta_{x-\mu,\mu}
     U^*_{x-\mu,\mu}\right]\gamma_\mu\; .
\ee
The effective action on the interface is
\be \label{7.27}
e^{S_i(U)} = \frac{_U<0-|0+>_U}{_1<0-|0+>_1}
             e^{i\left[\Phi_{+}(U)-\Phi_{-}(U)\right]}\; .
\ee
The $|0\pm >_U$ are the ground states of the Hamiltonian in presence of the
gauge fields whereas $|0\pm>_1$ denote the ground states for
free fermions. The phases $\Phi_\pm(U)$ are undetermined
which again reflects the phase ambiguity we have encountered
in (\ref{eq7.17},\ref{eq7.18}). The phases are fixed using the same condition
as in perturbation theory also for the non-perturbative case 
\be
e^{i\Phi_\pm(U)} = \frac{_U<0\pm |0\pm >_1}{|_U<0\pm |0\pm >_1|}\; .
\ee
In close analogy to the earlier discussion of the anomaly an external gauge 
field is chosen
\be
U(k) = e^{i\frac{A_\mu(\phi)}{L}\cos (\frac{2\pi k}{L}(n+\frac{1}{2}) )}
\ee
with
\be
A_\mu(\phi) = A_\mu +2\phi\sin(\frac{\pi k}{L})\; .
\ee
The anomaly equation now reads
\be
-\frac{i}{\pi}\lim_{L\rightarrow\infty}\frac{\partial S(A_\mu(\phi))}
             {\partial\phi}|_{\phi=0}
          = -\frac{1}{4\pi}\left[A_2k_1\right]\; .
\ee
In order to compute the anomaly numerically, the Dirac sea has to be filled
and the eigenfunctions corresponding to the negative energy eigenstates
have to be found. This amounts to find the eigenvalues of 
(\ref{eq7.24}) numerically.
In the computation 
$A_2$ was taken to be $A_2=0.32$.
In \cite{nane} two values of the domain wall mass were chosen, $m=0.5$ and
$m=0.9$ and the anomaly was computed on different size lattices. At 
both $m$ the value of the anomaly was extrapolated to $L=\infty$. Both
extrapolated values coincide and give $-0.02545(5)$. This is to be compared
to
the value of the anomaly in the continuum of $-0.02546$. 
The same test of the anomaly equation to be fulfilled on the finite lattice 
can also be performed in four dimensions by a clever choice of the
external gauge fields keeping the Hamiltonian to be at
least block diagonal. Again the continuum value of the strength of the anomaly 
could be reproduced in the thermodynamic limit. 

We discussed in this section a proposal of keeping  
the extra dimension in the domain wall model
strictly infinite. Alternatively we might think of having
introduced infinitely many regulator fields
for a chiral gauge theory.
We saw, that in continuum perturbation theory the correct value of the
consistent anomaly
is found. Even more remarkable is that the overlap formula resulting from
the ground state projection can be implemented on the finite lattice. The task
there is to diagonalize a --admittedly huge-- matrix. Choosing 
external gauge fields that vary smoothly, the value of the consistent
anomaly could be reproduced when the finite lattice results are
extrapolated to the infinite lattice. With suitable techniques of
matrix diagonalization a numerical simulation may be reached eventually.

With this outlook it might be possible to test whether a chiral gauge theory
can be obtained when the gauge fields are made dynamical. Of course,
the tests being performed on the overlap formula so far
correspond to the situation in the domain wall model where everything worked.
There we also found for smooth external gauge fields that the correct
--in this case covariant anomaly-- can be reproduced on the lattice
and that the chiral zeromode spectrum comes out as desired. This encouraging
picture was only destroyed when dynamics had been added.
Therefore it seems that only a full simulation can decide the question
whether the infinite $s$ approach is the solution of the chiral fermion
problem on the lattice.

In a numerical simulation only the real part would be used for the updating. 
Given a new gauge field configuration consisting of a change of only one
link, the imaginary part would be computed by diagonalizing $H_\pm$. It is
then hoped that, since only one link is changed at a time, the old eigenfunction
of $H_\pm$ is a good starting point for the new diagonalization process.
The general idea is not to take the imaginary part in the updating process
but into the measured observables. It is, of course, not clear at all
that the fluctuations of the imaginary part for realistic lattice gauge 
field configurations are small. If these fluctuations are large,
which might well
be possible due to the roughness of the 4-dimensional gauge fields, this
would not lead to any practical simulation algorithm. 
On the other hand an implementation of
the above simulation proposal 
could provide the first measure for the
strength of the fluctuations of the imaginary part of chiral fermion actions.

One might find the idea unsatisfactory 
that a chiral theory can only be constructed by
considering the Hamiltonians for both sides of the wall and construct
the overlap of the corresponding ground states.
There seems to be at the moment no way of connecting the overlap setup
with any euclidean lattice formulation of the same theory. 
However, 
the anomaly in the infinite dimensional lattice approach
for chiral lattice fermions {\em can} be understood from a limiting
procedure of a lattice version with finite extend in the $s$-direction
\cite{shamir93.2}. In this version the $s$-direction is split into
a gauged region for $s < N$ and an ungauged region for 
$N < s^{'} < N^{'} <\infty$. 
The limiting procedure which gives the correct --consistent-- anomaly
is to let both $N$ and $N'$ go to infinity simultaneously. Therefore
the interface effective action (\ref{eq7.3}) might be understood 
as the limit $s\rightarrow\infty$
of a gauge variant finite lattice version. 
As we have seen in the discussion
of the waveguide model, section~5.2, a gauge variant action can be made 
gauge invariant by adding scalar or St\"uckelberg fields. But then 
the infinite $s$ effective action is hiding that it came from the
waveguide model on the finite lattice. As the waveguide model
failed in constructing a chiral gauge theory, the same faith may
await the overlap approach to chiral fermions on the lattice.

\section{Domain wall fermions and QCD}

Even if the domain wall fermion approach to chiral gauge theories on the
lattice may fail, there could be a quite unexpected application of this idea.
As put forward by Shamir \cite{shamir93,shamir94} domain wall fermions could
be used for simulating QCD. 
This seems to be a natural approach since
the odd dimensional theory one starts with is 
vectorlike. 
Consider a slightly modified action for
domain wall fermions where we choose $2N$ points in the $s$-direction.
\bea \label{eq8.1}
  S_{F} & = & \sum_{x,s} \sum_{\mu =1}^4
\psibar_{x}^s\left[(\dslash(U) -\Delta(U) 
                          + m\right]\psi_{x}^s \nonumber \\
    & + &  \frac{1}{2}\sum_{x,s}\psibar_{x}^s\left[
           \delta_{s,s+\mu_5}(1+\gamma_5) 
          + \delta_{s-\mu_5,s}(1-\gamma_5) - 2\delta{s,s}\right]\psi_x^s
\nonumber \\
   &  + & \overline{m}\psibar_x^{2N}(1+\gamma_5)\psi_x^1 
          + \overline{m}\psi_x^1(1-\gamma_5)\psi_x^{2N}\; .
\eea
This action is identical to eq.(\ref{eq7.1}) except that the boundary term
in the $s$-direction is made explicit. 
The gauge fields are kept 4-dimensional.
The boundary terms are equipped with an additional new parameter $\overline{m}$.
For $\overline{m}=1$, which corresponds to periodic boundary conditions,
no zeromodes appear. For $\overline{m}=0$ 
we will have open boundary conditions and the action (\ref{eq8.1}) 
describes boundary fermions with chiral surface modes \cite{shamir93}.

For a non-zero value of $\overline{m}$ an interaction of the surface modes
at $s=1$ and $s=2N$ is induced and they will combine to a Dirac state.
In order to derive the mass dependence of this state we 
compute the propagator from the action (\ref{eq8.1}). First there is
a homogeneous solution of the Dirac equation
\be \label{eq8.2}
G_0(s,s') = Be^{-\alpha|s-s'|}
\ee
with 
\bea \label{eq8.3}
B^{-1} & =& 2 b\sinh\alpha \nonumber \\
2\cosh\alpha(p) & = & \frac{1+b^2(p)+\overline{p}^2}{b(p)} \nonumber \\
\overline{p}=\sin p_\mu & & b(p) = 1-m+\sum_\mu(1-\cos(p_\mu))\; .
\eea
The full propagator is given as
\bea \label{eq8.4}
G(s,s') & = & G_0(s,s') + A_{-}e^{-\alpha(s+s')}+A_{+}e^{-\alpha(2N-s-s')} 
\nonumber \\
        & + & A_m\left( e^{-\alpha(N+s-s')} + e^{-\alpha(N+s'-s)}\right)
\eea
where
\bea \label{eq8.5}
A_{-} & = & \Delta^{-1} Bb(1-\overline{m}^2)(e^\alpha -b) \nonumber \\
A_{+} & = & \Delta^{-1} Bb(1-\overline{m}^2)(e^{-\alpha} -b) \nonumber \\
A_m & = & 2\Delta^{-1}Bb^2\overline{m}\cosh\alpha \nonumber \\
\Delta & = & be^\alpha-1+\overline{m}^2b(e^\alpha-b)\; .
\eea

This complicated expression can be evaluated in the limit of small 
momenta and small values of $\overline{m}$. As a result one finds that the
propagator in momentum space is
\be \label{eq8.7}
G(p^2) = p^2 + \overline{m}^2 m^2 (2-m)^2\; ;\;\; 
                   p^2\; ,\overline{m}^2 \ll 1\; .
\ee
For the boundary fermion model $\overline{m}=0$. We see that in the low momentum
limit the propagator describes a massless 
particle. The $s$-dependence of the full propagator (\ref{eq8.4}) shows
that the chiral zeromodes appear as surface modes on the boundary of
the lattice in the $s$-direction. This is the same result as we discussed
in section 2 where the zeromodes were computed in the Hamiltonian language.
They are depicted in figure 1b. 

Turning on $\overline{m}$, these zeromodes are coupled through the
links connecting the last and the first $s$-slice and acquire a --Dirac--
mass. As can be read off from (\ref{eq8.7}) this mass is proportional to
$\overline{m}$ which might therefore be interpreted as 
an external quark mass. The reason,
why this result makes boundary fermions interesting for QCD simulations is the
following. According to the nogo theorems chiral symmetry is broken on the
lattice. However, as weak coupling perturbation theory shows, chiral
symmetry will be restored in the continuum limit of lattice QCD. To reach this
limit a careful tuning of the bare parameters in the lattice QCD action
is necessary to approach the phase transition. The tuning of the pion mass
to zero as a function of the bare parameters is non-trivial and demanding.
In particular, there are $O(1/a)$ corrections to the renormalized quark
mass.

Eq.(\ref{eq8.7}) offers a nice alternative. 
It suggests that quark masses only get multiplicative renormalization and
that one has therefore 
only to tune $\overline{m}$ to zero in order to reach the phase transition. 
In this context, $\overline{m}$ might be thought of as an external source.
In
particular, the phase transition point is expected to be $\overline{m}=0$.
For a finite lattice in the $s$-direction there is an additional
complication.
Since the wavefunctions of both modes on the walls fall off exponentially,
there will be always an overlap of both quark modes in the middle of the 
lattice. The value of the quark's wavefunction can be estimated to be
$(1-m)^s$.  
Therefore the extension of the lattice in the extra dimension
should be $N\approx -log(m)$. 
Whether the advantage to make tuning 
trivial will merit the introduction of an extra dimension can only be
answered in actual simulations.

For a finite system the overlap of the surface modes lead to some
anomalous effects in the middle of the lattice. The currents are defined as
\be
j_\mu =\frac{1}{2}\left(\psibar_x^s(1+\gamma_\mu)U_{x,\mu}\psi_{x+\mu}^s
-\psibar_{x+\mu}^s(1-\gamma_\mu)U_{x-\mu,\mu}^\dagger\psi_{x}^s\right)
\ee
for $\mu=1,...,4$ and
\be 
j_5 =   \left\{  \begin{array}{cc}
\psibar_x^sP_R\psi_{x}^{s+1} - \psibar_x^{s+1}P_L\psi_{x}^{s} & 
                                                               1\le s < 2N \\

\psibar_x^{2N}P_R\psi_{x}^{1} - \psibar_x^{1}P_L\psi_{x}^{2N} & 
                                                                s = 2N 

               \end{array} \right. \; .
\ee
Since the 5-dimensional theory is vectorlike, it is anomaly free and
the divergence equation is $\partial_\mu j_\mu=-\partial_5j_5$, $\mu=1,...,4$.
Written out explicitly 
\be
\sum_{\mu=1}^4 \partial_\mu j_\mu = \left\{ \begin{array}{cc}
       -j_5^{s=1} -\overline{m}j_5^{s=2N} & s=1 \\
       -\partial_5j_5^s                & 1 < s < 2N \\
        -j_5^{s=2N-1} -\overline{m}j_5^{s=2N} & s=2N
       \end{array} \right. \; .
\ee
One can define an axial current as in section~3 
\be 
A_\mu = -\sum_{s=1}^{2N}\mbox{sign}(N-s+\frac{1}{2})j_\mu\; .
\ee
Then the divergence is
\be
\partial_\mu A_\mu = 2\overline{m}j_5^{2N} + 2 j_5^{N}\; .
\ee
Whereas $j_5^{N}$ appears to be derived from a usual classical
massterm, $j_5^{2N}$ is an additional anomalous term
stemming from the overlap of the surface modes. As is shown in
\cite{shamir94} this term vanishes with $N\rightarrow\infty$ as

\be
\exp(-c_0/g_0^2)\exp(-\Lambda N)
\ee
with $c_0$ a constant, $g_0$ the gauge coupling and $\Lambda$ 
the confinement length.
From this formula it is seen that for every $g_0$ fixed, the anomalous term
vanishes if the extent of the lattice in the extra dimension is sent to 
infinity. This suggests that one might try to follow the program of the previous
section and work in the $N=\infty$ limit directly. 
Since in the application of the domain wall model to QCD the effective
action on the boundary is real and therefore does not possess the
phase ambiguity we found for the chiral gauge theory case, the program
of Narayanan and Neuberger can now be performed without subtleties
involved. The derivation of the overlap formula for the QCD case can be 
worked out following \cite{nane} closely and an explicit formula is
found in \cite{shamir94}.
The vanishing of the anomalous term in the $N\rightarrow\infty$ limit
constitutes a proof that axial symmetry is restored at the QCD
phase transition. In the derivation of this result basically no
approximation is involved. This amounts to the statement that
restoration of axial symmetry will also appear at strong coupling and is
valid not only to every order in weak coupling perturbation theory 
\cite{kasmi} but holds also non-perturbatively.
 
\section{Summary and conclusions}

In this article we summarized the status of domain wall fermions 
invented to construct a chiral gauge theory on the lattice. 
In the continuum domain wall fermions are 
obtained as chiral zeromodes bound to a
--soliton shaped--
mass defect living in an extra odd dimension. The domain wall model can be
taken over to the finite lattice by choosing the mass defect to
be a step function. Through the necessary introduction
of boundary conditions a second anti-wall is generated and
the chiral zeromodes appear pairwise as left and
righthanded fermions. They are bound to the wall and the anti-wall, respectively,
and 
fall off exponentially with increasing distance from the walls. 
Alternatively one may choose free boundary conditions
in the extra dimension 
in which case
the chiral modes appear as surface modes.

The charge flow in the domain wall model is according to the picture
developed in the continuum by Callan and Harvey. The 
anomaly appearing in the
zeromode current along the wall is compensated by the divergence
of the Chern-Simons current induced by the heavy fermions that live off
the wall in the extra dimension. 
The Callan-Harvey 
analysis can be performed on the lattice confirming the continuum 
picture at least,
as long as the gauge fields are taken to be smooth and external.
The correct strength of the --covariant-- anomaly can be directly measured
on the finite lattice. 

The crucial step for the realization of a chiral gauge theory on the lattice
is the coupling to dynamical gauge fields. The fluctuations of the gauge field
can become strong and the fear is that they will lead
to an interaction of the two chiral zeromodes combining them to a 
Dirac particle. 
In the waveguide model only one of the two
domain walls is gauged. Interpreting the extra dimension as a sophisticated
flavour space, the gauge fields are kept strictly 4-dimensional and the same
for all flavours in a region around the ``wall''-flavour, i.e. the location
of the wall. At the boundaries of this ``waveguide region'' gauge invariance
is lost. This can be repaired by introducing scalar or St\"uckelberg
fields at the boundary of the waveguide region. The coupling of the scalar 
fields and the fermions are of the Yukawa-form and will be equipped with a
Yukawa-coupling $y$. For $y=0$ one essentially obtains the model with free
boundary conditions and new fermionic zeromodes appear which can interact
with the domain wall zeromode through the gauge fields. 
The important and crucial question for the model to succeed is then whether
the fermionic spectrum at the waveguide boundary can be made massive
such that the modes decouple in the continuum limit and one is left
with only the chiral zeromode on the gauged wall.

The result of the investigations in the domain wall model can be given,
comparing fig.9 and fig.13. These figures show the phase diagrams that
would have led to a possible construction of a chiral gauge theory
with domain wall fermions (fig.9) and the much more realistic phase
diagram 
\footnote{In Fig.13 the PMS$_2$ phase is left out. The existence of
this phase was not investigated. But even if it would exist, 
a construction of a chiral gauge theory would not be possible because
all modes including the one on the domain wall become heavy.} 
in fig.13
which prohibit such a construction. The phase diagram
in fig.13 was obtained by a large-$y$ expansion in combination with
numerical simulation results. 

The interesting part of the phase diagram in fig.9 is the PMS$_1$ phase.
In this phase the fermions at the boundary of the waveguide region
would have become massive. There the Yukawa-coupling would have been
strong enough to bind the fermion and the scalar field to form
a massive bound state. Therefore this state would have decoupled from the
low energy spectrum, leaving behind a chiral zeromode on a domain wall
coupled to gauge fields in a gauge invariant manner. 
In earlier investigations of coupled fermion-Higgs models such a phase
could be established or it was missing, depending on the particular 
implementation of the form of the Yukawa-coupling. 
What is found in the waveguide model from the numerical simulations
and the large-$y$ expansion is that the PMS$_1$ phase is absent. Instead,
there are two symmetric phases with weak coupling behaviour both of them with a 
{\em vectorlike}  spectrum of zeromodes. Thus in neither of these
symmetric phases a chiral gauge theory can be obtained. A word of care
has to be given at this point. Although the phase diagram in fig.13 represents
a consistent picture from the large $y$ expansion and numerical simulations,
the simulations were very difficult in the most interesting region of the
phase diagram. Although evidences are strong that a PMS$_1$ phase is absent,
its non-existence could not be proven without any doubt. Maybe using
the algorithm proposed in \cite{boson} can help to achieve better simulation
results in the difficult region.

\begin{figure}                                                              
\vspace{10.0cm}
\includegraphics{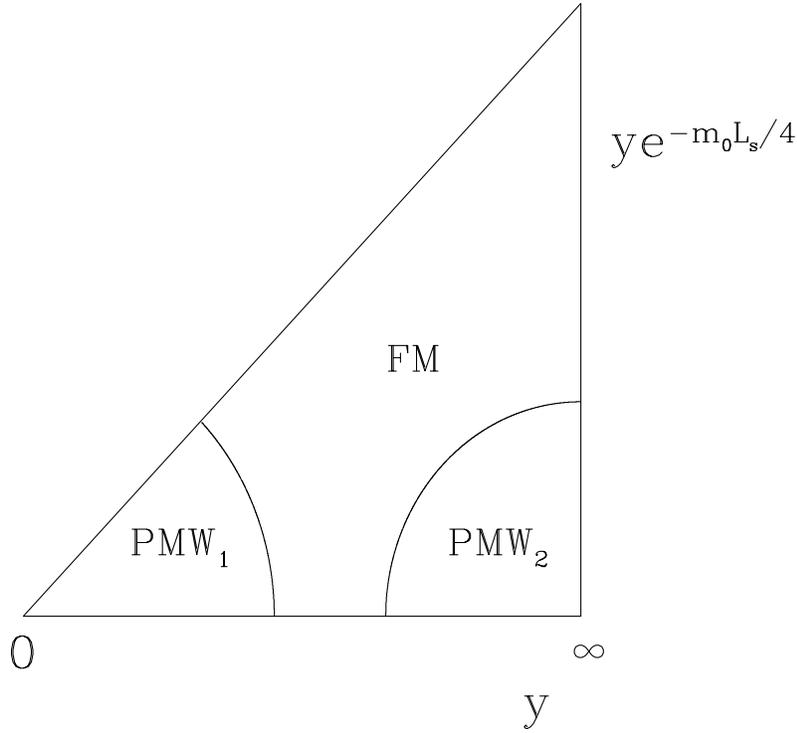}
                                                                             
\caption{ \label{fig13}                                                         
A realistic phase diagram for domain wall fermions.
There will be a symmetry broken (FM) phase as well as 
two symmetric phases (PMW$_1$ and PMW$_2$) with different vectorlike spectra
of massless fermions. 
}                                                                              
\end{figure}                                                                   

What remains from the domain wall model idea is a variant where the extra
dimension is kept strictly infinite. In this interesting line of thinking
the extra dimension is regarded as the time axis of a 4-dimensional 
Hamiltonian. The Hamiltonian is split into two parts corresponding to the two
sides of the domain wall. Due to the infinite extension in the extra dimension
the ground states on both sides of the wall are projected out. The resulting
theory consists of the overlap of these two ground states. Assuming smooth
external gauge fields it could be shown in perturbation theory
and non-perturbatively through a diagonalization of the finite
lattice Hamiltonians that this formulation of the
domain wall model gives again the correct --consistent-- anomalies. 
So far, in this interesting approach no
dynamical gauge fields have been considered and it remains to be seen
whether the promising results survive their inclusion.

A completely different application results from the interpretation
of the wall zeromodes as massless quarks. 
The boundaries of the lattice in the extra dimension 
can be coupled through a mass term.
In this way, the surface modes combine to build
a Dirac fermion. 
The interesting aspect of this approach is that the mass of the
Dirac-particle is directly proportional to the new mass parameter. Thinking
of simulations in QCD this would mean that one could perhaps avoid the
fine tuning problem of the quark mass in the conventional formulations
of lattice QCD.

To conclude, the clash between chiral gauge invariance and the existence
of a single Weyl fermion on the lattice could also not be resolved
by the domain wall fermion approach. It seems that non-perturbative lattice
models of chiral gauge theories unavoidingly lead
to doubler states which might be interpreted as
mirror fermions. 
One is therefore left with the choice to believe in some --yet undiscovered--
new fermions, or to think further,
maybe along the lines of the Rome-approach \cite{romeapproach}, 
how nature managed to construct
chiral gauge theories.

\section*{Acknowledgement}

This work would not have been possible without the very enjoyable
collaboration with M.F.L. Golterman, D.B. Kaplan, D.N. Petcher, M. Schmaltz and
J.C. Vink. I am grateful to them for numerous discussions and suggestions.
In particular I would like to thank M.F.L. Golterman for reading the 
manusscript.





\end{document}